\documentclass[a4paper,numbook]{svjour3}
\usepackage{amssymb,amsmath,amsfonts,latexsym}
\usepackage{epsfig}
\usepackage{times}
\usepackage[english]{babel}
\usepackage[T1]{fontenc}
\usepackage[latin1]{inputenc}

\begin{document}

\title{A Motivating Exploration on Lunar Craters and Low-Energy Dynamics in the Earth -- Moon System}

\titlerunning{Low-energy lunar impacts}       

\author{Elisa Maria Alessi \and  Gerard G\'omez \and Josep J. Masdemont}

\authorrunning{Alessi, G\'omez, Masdemont} 

\institute{E.M. Alessi \at
           IEEC \& Dpt. Matem\`atica Aplicada i An\`alisi, Universitat de Barcelona \\
           Gran Via 585, 08007 Barcelona, Spain.\\
           \email{elim@maia.ub.es}           
           \and
           G. G\'omez \at
           IEEC \& Dpt. Matem\`atica Aplicada i An\`alisi, Universitat de Barcelona \\
           Gran Via 585, 08007 Barcelona, Spain.\\
           \email{gerard@maia.ub.es}  
           \and
           J.J. Masdemont \at
           IEEC \& Dpt. Matem\`atica Aplicada I, Universitat Polit\`ecnica de Catalunya \\
           Diagonal 647, 08028 Barcelona, Spain.\\
           \email{josep@barquins.upc.edu}  
}

\date{Received: date / Accepted: date}

\maketitle

\begin{abstract}
\noindent 

It is known that most of the craters on the surface of the Moon were created by the
collision of minor bodies of the Solar System. Main Belt Asteroids, which can
approach the terrestrial planets as a consequence of different types of resonance,
are actually the main responsible for this phenomenon.

Our aim is to investigate the impact distributions on the lunar surface that
low-energy dynamics can provide. As a first approximation, we exploit the
hyberbolic invariant manifolds associated with the central invariant manifold around
the equilibrium point $L_2$ of the
Earth -- Moon system within the framework of the Circular Restricted Three -- Body
Problem. Taking transit trajectories at several energy levels, we look for
orbits intersecting the surface of the Moon and we attempt to define a
relationship between longitude and latitude of arrival and lunar craters
density. Then, we add the gravitational effect of the Sun by considering
the  Bicircular Restricted Four -- Body Problem. 

In the former case, as main outcome we observe a more
relevant bombardment at the apex of the lunar surface, and a percentage of impact
which is almost constant and whose value depends on the Earth -- Moon distance
$d_{EM}$ assumed. In the latter, it seems that the Earth -- Moon and Earth -- Moon
-- Sun relative distances and the initial phase of the Sun $\theta_0$ play a crucial
role on the impact distribution. The leading side focusing becomes more and more
evident as $d_{EM}$ decreases and there seem to exist values of $\theta_0$ more
favorable to produce impacts with the Moon. Moreover, the presence of the Sun make
some trajectories to collide with the Earth. The corresponding percentage floats
between 1 and 5 $\%$.

As further exploration, we assume an uniform density of impact on the lunar surface,
looking for the regions in the Earth -- Moon neighbourhood these colliding
trajectories have to come from. It turns out that low-energy ejecta originated from
high-energy impacts are also responsible of the phenomenon we are considering.


\end{abstract}

\section*{REMARK}

The paper is being published in Celestial Mechanics
and Dynamical Astronomy, vol. 107 (2010).

\section{Introduction}\label{sec:1}

The surface of the Moon is constellated by impact craters of various sizes, mainly generated from the collision of objects coming from the Main Asteroid Belt. Indeed, the Inner Solar System can be reached by such minor bodies as a consequence of different types of resonance \cite{BMJPLMM}. The intense lunar bombardment took place between 3.8 and 4 Gy ago, being at the present day the meteroidal flux about $10^3$ lower \cite{H}.

The cratering process is interesting for several branches of science. First of all, by comparing densities of craters on different surfaces it is possible to derive the relative age of the corresponding terrains (see, for instance, \cite{NIH,SR,MMCMM}). Roughly speaking, the greater the density the older the surface. Also, the geological chronology of the terrestrial planets is now becoming more and more accurate thanks to the space missions that provide radiometric age estimates for different regions. This is especially true if we take as reference case the Moon, for which a great amount of data is now available. From this kind of analysis, new insight on the Solar System evolution can be obtained. As further aspect, the flux of impacts offers information on the Solar System minor bodies population. 

The main problem in all these studies resides on the fact that the crater formation is a phenomenon not fully understood yet. There does not exist a predictive, quantitative model of crater formation, that is, a reliable methodology that can be applied to all situations. The size of the crater that forms at the end of the excavation stage depends on the asteroid's size, speed and composition, on the collision angle, on the material and structure of the surface in which the crater forms and on the surface gravity of the target \cite{M}. The problem in the determination of the crater's dimension concerns with the poorness of the experimental or observational data. This difficulty is usually overcome by extrapolating beyond experimental knowledge through scaling laws.

This work regards the paths that impacting asteroids might have followed. In particular, we will deal with low-energy trajectories, first derived in the Circular Restricted Three -- Body Problem (CR3BP) framework applied to the Earth -- Moon system and then analysed also accounting for the Sun gravitational attraction by means of the Bicircular Restricted Four -- Body Problem (BR4BP). We assume the minor bodies to have already left the Main Asteroid Belt and we consider as main entrance to the Earth -- Moon neighbourhood the stable invariant manifold associated with the central invariant manifold corresponding to the $L_2$ equilibrium point. 

We will look for the distribution of impacts that such orbits can create, paying
attention to the fact that the Moon is locked in a $1:1$ spin--orbit resonance.
In particular, we wonder if, for the range of energy under consideration, the
Moon acts as a shield for the Earth or if the greatest concentration of
collisions still takes place on the leading side of the surface, as other
authors have pointed out with different approaches. See, for example,
\cite{HN,MF,LW}.

In a second step, from a backward integration, we attempt at discovering any other gate that can lead to a lunar impact within low-energy regimes.

We recall that due to the small values of energy we consider, the impacts obtained can yield to at most $40$ km in diameter craters. This value has been computed by applying the scaling laws of Melosh \cite{M} to the Moon's surface with an impact velocity corresponding to the escape lunar velocity (about $2.4$ km$/$s).

\section{The Circular Restricted Three -- Body Problem}\label{sec:2}         
                                                                
The Circular Restricted Three -- Body Problem \cite{S} studies the
behaviour of a particle $P$ with infinitesimal mass $m_3$ moving
under the gravitational attraction of two primaries $P_1$ and $P_2$,
of masses $m_1$ and $m_2$, revolving around their center of mass
in circular orbits.                                             
                                                                
To remove time from the equations of motion, it is convenient to introduce a
synodical reference system $\{O,x,y,z\}$, which rotates around the $z$--axis
with a constant angular velocity $\omega$ equal to the mean motion $n$ of
the primaries. The origin of the reference frame is set at the barycenter
of the system and the $x$--axis on the line which joins the primaries,
oriented in the direction of the largest primary. In this way we work with
$m_1$ and $m_2$ fixed on the $x$--axis.                         
                                                                
The units are chosen in such a way that the distance between the primaries and the modulus of
the angular velocity of the rotating frame are unitary. We define the mass ratio $\mu$ as $\mu = \frac{m_{2}}{m_{1}+m_{2}}.$ For the Earth -- Moon
system $\mu = 0.012150582$.                                                   
                                                                
The equations of motion for $P$ can be written as                                     
\begin{eqnarray}
\ddot x-2\dot y&=& ~x-\frac{(1-\mu)}{r_{1}^{3}}(x-\mu)-\frac{\mu }{r_{2}^{3}}(x+1-\mu),
\nonumber \\
\label{eq:1}
\ddot y+2\dot x&=& ~y-\frac{(1-\mu)}{r_{1}^{3}}y-\frac{\mu }{r_{2}^{3}}y,\\
\ddot z&=&~-\frac{(1-\mu)}{r_{1}^{3}}z-\frac{\mu}{r_{2}^{3}}z,\nonumber
\end{eqnarray}
where $r_{1}= [(x-\mu)^{2}+y^{2}+z^{2}]^{\frac{1}{2}}$ and
$r_{2}=[(x+1-\mu)^{2}+y^{2}+z^{2}]^{\frac{1}{2}}$ are the distances from
$P$ to $P_{1}$ and $P_{2}$, respectively.

The system (\ref{eq:1}) has a first integral, the Jacobi
integral, which is given by
\begin{equation}\label{eq:2}
C=x^{2}+y^{2}+{\frac{2(1-\mu)}{r_{1}}}+{\frac{2\mu}{r_{2}}}+(1-\mu)\mu -\left( \dot{x}^{2}+\dot{y}^{2}+\dot{z}^{2}\right).
\end{equation}
In the synodical reference system, there exist five equilibrium or \emph{libration}
points. Three of them, the collinear ones,
are in the line joining the primaries and are usually denoted by
$L_1$, $L_2$ and $L_3$. If $x_{L_i}$ ($i=1,2,3$) denotes
the abscissa of the three collinear points, we assume that
$$x_{L_{2}}<\mu-1<x_{L_{1}}<\mu<x_{L_{3}}.$$
Let $C_i$ be the value of the Jacobi constant at the $L_i$ 
equilibrium point.  We have the following relation
\begin{eqnarray}\nonumber
C_1>C_2>C_3>C_4=C_5.
\end{eqnarray}

\begin{figure}[t!]
\centering
\begin{tabular}{cccc}
\includegraphics[width=35mm]{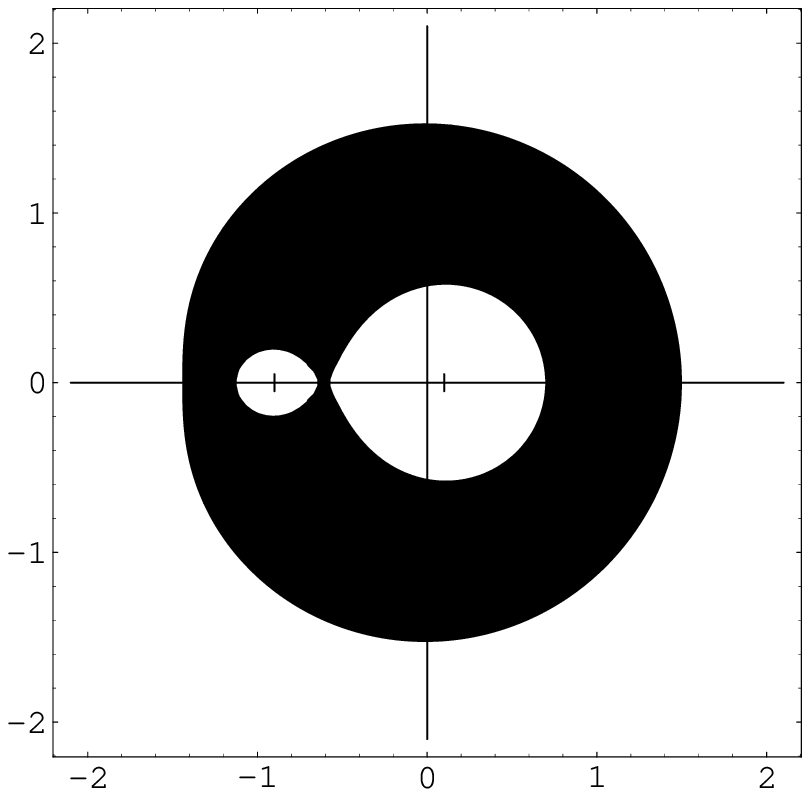} &
\includegraphics[width=35mm]{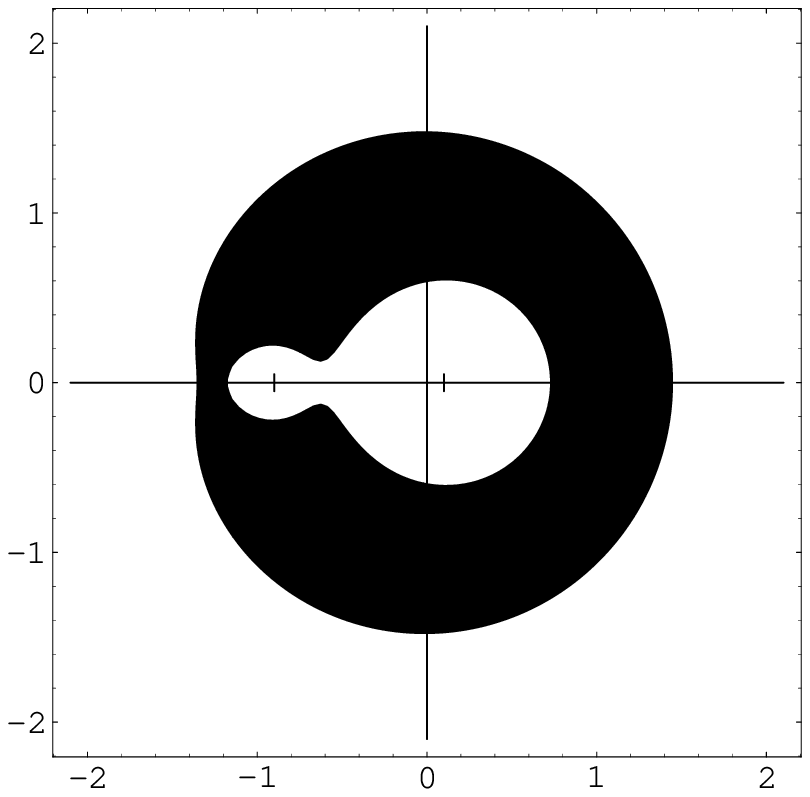} &   
\includegraphics[width=35mm]{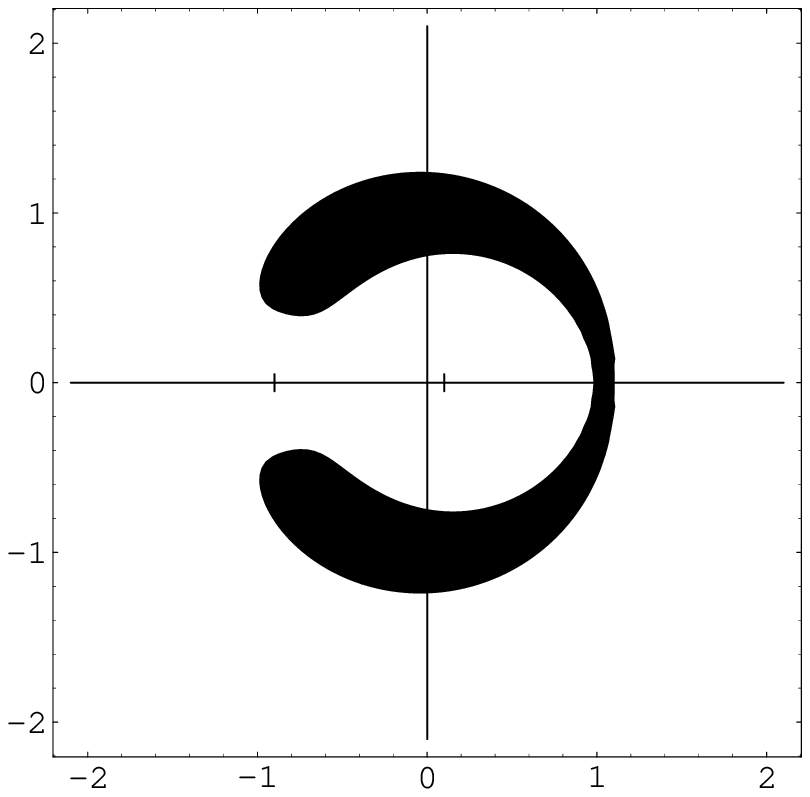} &
\includegraphics[width=35mm]{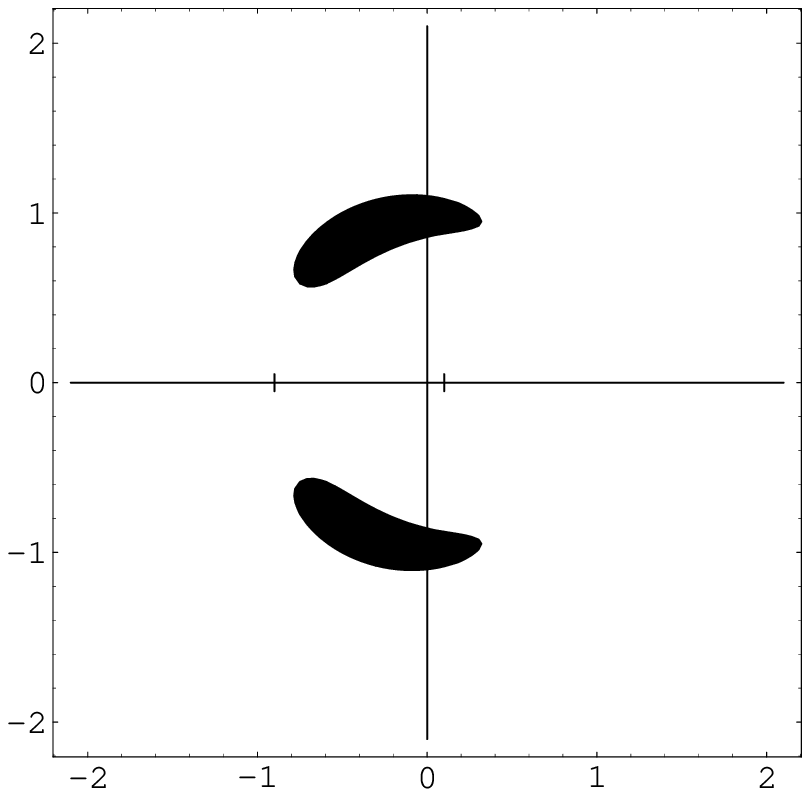} \\
$C>C_1$  & $C_2<C<C_1$ &    $C_3<C<C_2$ & $C_4<C<C_3$
\end{tabular}
\caption{\sf Zero-velocity curves (the intersection of the zero-velocity
surfaces with the $\{z=0\}$ plane) for $\mu>0$. The motion is forbidden in the
filled areas. The tick marks on the horizontal axis show the position of the
primaries: $P_1$ on the positive $x$--axis and $P_2$ on the negative $x$--axis. The case $C<C_4=C_5$ is not displayed since the motion is allowed everywhere.}
\label{fig:1}
\end{figure}

Depending on the value of the Jacobi constant, it is possible to define where
the particle can move in the configuration space. These regions are known
as Hill's regions and their boundaries are the zero-velocity
surfaces. For a given mass parameter, there exist five
different geometric configurations, corresponding to five different energy 
levels.  
If $C>C_1$, the infinitesimal mass can just move either in a neighbourhood
around the largest primary or in a small neighbourhood around the smallest
one. If $C_2<C<C_1$, it can move from the neighbourhood of one primary to the
neighbourhood of the other one. If $C_3<C<C_2$, it occurs the so--called
\emph{bottleneck} configuration, that is, the accessible region opens up
beyond $m_2$. On the other hand, $P$ can go toward $L_1$ and $L_3$ when
$C_4<C<C_3$.  Finally, if $C<C_4$ there are not forbidden regions. In
Fig.~\ref{fig:1}, we represent the intersections of the zero-velocity
surfaces with the $\{z=0\}$ plane. These intersections are known as zero-velocity curves. 

The collinear libration points behave as the product of two
centers by a saddle.  This means that around a collinear point we deal with bounded orbits, which are due to the central part and also with escape trajectories, which depart exponentially from the neighbourhood of the collinear point for $t\rightarrow\pm\infty$ and are due to the saddle component. The former kind of motion belongs to the {\em central invariant manifold}, the latter to the {\em hyperbolic invariant manifolds} associated with the central invariant one. The hyperbolic manifolds consist, in particular, in one stable and one unstable.

To be more precise, when we consider all the energy levels, the center $\times$ center part
gives rise to 4--dimensional central manifolds around the collinear equilibrium points \cite{GJMS}. Different types of periodic and quasi-periodic orbits fill the central invariant manifold: we refer to them as {\em central orbits}. On the other hand, due to the hyperbolic character, the dynamics close to the collinear libration points is that of an
unstable equilibrium. This means that each type of central orbits around $L_1$, $L_2$ and $L_3$ has a stable and an
unstable invariant manifold. Each manifold has two branches, a positive and a negative
one. They look like tubes of asymptotic trajectories tending to, or departing from, the corresponding orbit. These tubes have a key role
in the study of the natural dynamics of the libration regions. When going forwards in time, the trajectories on the stable
manifold approach exponentially the orbit, while those on the unstable manifold depart exponentially.  As a matter of fact \cite{LMS,GKLMMR}, these orbits separate two types of motion. The {\em transit} solutions are those orbits belonging to the interior of the manifold and passing from one region to another. The {\em non-transit} ones are those staying outside the tube and bouncing back to their departure region.

The main idea of this work is the lunar impact trajectories to be driven by the stable component associated with each type of central orbit. We actually adopt a methodology that does not need to compute every family of central orbit, but for sake of completeness we describe the most exploited ones. Planar and vertical Lyapunov orbits, halo orbits and Lissajous orbits are all central orbits and underlie the approach and the results of this study. 

\begin{figure}[h!] 
\begin{center}
\includegraphics[width=100mm]{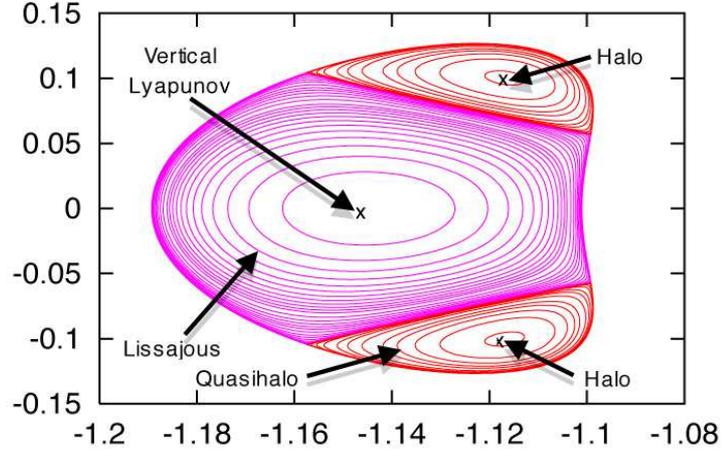}
\caption{\sf We show the Poincar\'e section at $\{z=0\}$ corresponding to the $L_2$
point of the Earth -- Moon system for $C=3.142003$. In this $(x,y)$ projection we can distinguish clearly different types of periodic and quasi-periodic orbits, all  belonging to the central invariant manifold associated with $L_2$ and thus denoted as central orbits.}
\label{fig:2_new}
\end{center}
\end{figure}

According to \emph{Lyapunov's centre theorem}, each centre gives rise\footnote{This
statement is true unless one of frequencies is an integer multiple of the
other, which happens only for a countable set of values of mass ratio.} to a
family of periodic orbits, whose period tends respectively to the frequencies related to both centers, $\omega_p$ and $\omega_v$, when approaching the equilibrium point.  These families are known as \emph{vertical Lyapunov family} and \emph{planar Lyapunov
family}. The quasi-periodic
Lissajous orbits are those associated with two--dimensional tori, whose two basic
frequencies $\omega_1$ and $\omega_2$ tend, respectively, to the frequencies related to both centers, $\omega_v$ and $\omega_p$, when
the amplitudes tend to zero.  They are characterized by an harmonic motion in
the $\{z=0\}$ plane and an uncoupled oscillation in
$z$--direction with different periods. Furthermore, \emph{halo} orbits are 3--dimensional periodic orbits that show up at the first
bifurcation of the planar Lyapunov family. In fact, there appear two families
of halo orbits which are symmetrical with respect to the $\{z=0\}$
plane. They are known as {\sl north} and {\sl south class} halo families or
also \emph{first class} and \emph{second class} halo families. In Fig.~\ref{fig:2_new}, we give a representation of the above central orbits by means of the Poincar\'e section at $\{z=0\}$ for a given value of $C$.

Throughout the paper, we denote the central invariant manifold corresponding to a given collinear equilibrium point $L_i$ as $\mathcal{W}^c(L_i)$, and the stable and the unstable invariant manifold (the hyperbolic manifolds) associated with $\mathcal{W}^c(L_i)$ as $\mathcal{W}^s(L_i)$ and $\mathcal{W}^u(L_i)$, respectively.

\subsection{Transit orbits belonging to $\mathcal{W}^s(L_2)$}\label{sec:2.1}

As mentioned before, our aim is to study the role that low-energy orbits might have had in the formation of lunar impact craters. To this end, we assume as main channel to get to the Moon the stable invariant manifold associated with the central manifold around the $L_2$ point, $\mathcal{W}^s(L_2)$. This hypothesis is based on the fact that we admit as energy levels only those belonging to the third regime depicted in Fig.~\ref{fig:1}. In particular, we consider $C_3<C<C_2$, that is, $C\in(3.184163,3.024150)$ for the mass parameter under study. The reason for this choice is that under either the first or the second regime, there does not exist the possibility that a particle coming from the Outer Solar System collides with the Moon. On the other side, by discarding the more energetic ones we force the asteroids to approach the Moon before arriving to the Earth.

We need an efficient way to represent the dynamics driven by $\mathcal{W}^s(L_2)$ for each energy level, with no distinction on central orbits.
The main idea we develop is to determine $\mathcal{W}^s(L_2)$ using only the stable invariant manifolds of the planar and vertical periodic orbits.
In particular, for a well-defined value of $C$, we have
\begin{equation}\label{eq:3}
\mathcal{W}^s(L_2) \subset \mathcal{W}^s(Planar~Lyapunov) \times \mathcal{W}^s(Vertical~Lyapunov).
\end{equation} 
Transit trajectories of the stable invariant manifold associated with any central orbit lie inside the above product. 

In what follows, we do not explain how to compute these kinds of periodic orbits \cite{GJMS}, but how we derive the initial conditions on the associated hyperbolic manifold. Also, we give a hint on the derivation of (\ref{eq:3}).

\subsubsection{Numerical linear approximation}\label{sec:2.1.1}

There exist different methods to compute stable and unstable manifolds associated with
periodic orbits, here we consider the linear approximation, that makes use of the eigenvectors, corresponding to the hyperbolic directions, of the monodromy matrix of a given periodic orbit. The reader interested in other approaches should refer to \cite{MAS}.

Let $\mathbf{x}_{0}(t=0)$ be the initial condition of a $T$--periodic orbit, $\phi_t$ the flow at time $t$ under the CR3BP vectorfield, $D\phi_t$ its differential and
$M:=D\mathbf{\phi}_T(\mathbf{x}_{0})$ the monodromy matrix.  If the periodic
orbit is hyperbolic, then there exist $\lambda_j,\ \lambda_j^{-1} \in
\textrm{Spec} (M)$ such that $\lambda_j\in\mathbb{R}\slash\{-1,1\}$.  In this
case, there exist a stable and an unstable manifold, which are tangent,
respectively, to the $\lambda_j$ and $\lambda_j^{-1}$ eigendirections at
$\mathbf{x}_0(0)$.

Let $\mathbf{v}_S(0)$ and $\mathbf{v}_U(0)$ be, respectively, the normalized stable and unstable eigenvector corresponding to the point $\mathbf{x}_{0}(0)$
on the periodic orbit considered, being the normalization performed in such a way that $\mathbf{x}_{S,U}(0)$ has modulus equal to 1. We recall that in the CR3BP case, just one
hyperbolic eigenvector is sufficient to determine both
branches of both manifolds, since the stable and the unstable directions are
related by a symmetry relationship. More concretely, if $(v_1, v_2, v_3, v_4,v_5,v_6)$ is the eigenvector
associated with $\lambda_j$, then $(v_1,-v_2,v_3,-v_4,v_5,-v_6)$ is the
eigenvector associated with $\lambda_j^{-1}$.

The linear approximation for the initial conditions of the stable and the unstable manifold at $\mathbf{x}_{0}(0)$
is given, respectively, by
\begin{equation}
\begin{array}{rcl}
\mathbf{x}_S(0)&=&\mathbf{x}_{0}(0)\pm \epsilon\mathbf{v}_S(0),\\
\mathbf{x}_U(0)&=&\mathbf{x}_{0}(0)\pm \epsilon\mathbf{v}_U(0),
\end{array}
\label{eq:4}
\end{equation}
where $\epsilon$ is some small positive parameter. The value of $\epsilon$
fixes the size of the displacement we are performing from the periodic orbit to
the hyperbolic manifold, if we use the above mentioned normalization for $\mathbf{v}_S(0)$ and $\mathbf{v}_U(0)$. Its value must be chosen in such a way to guarantee
that $\mathbf{x}_{0}(0) \pm \epsilon\mathbf{v}_{S,U}(0)$ are still points where
the linear and nonlinear manifolds are close. However, it cannot be too small
to prevent from rounding errors. A typical value of $\epsilon=10^{-4}$ (in the
adimensional set of units) has been adopted in our computations. The sign of
$\epsilon$ determines the branch of the manifold.

If $t\ne 0$, we can exploit the following relations:
\begin{equation}
\begin{array}{rcl}
\mathbf{x}_S(t)&=&\phi_t(\mathbf{x}_{0}(0)) 
\pm \epsilon D\phi_t (\mathbf{v}_S(0)),\\
\mathbf{x}_U(t)&=&\phi_t(\mathbf{x}_{0}(0))
\pm \epsilon D\phi_t (\mathbf{v}_U(0)).
\end{array}
\label{eq:5}
\end{equation}

\begin{figure}[b!]
\begin{center}
\begin{tabular}{cc}
\hspace{-1.6cm}\includegraphics[width=60mm,angle=-90]{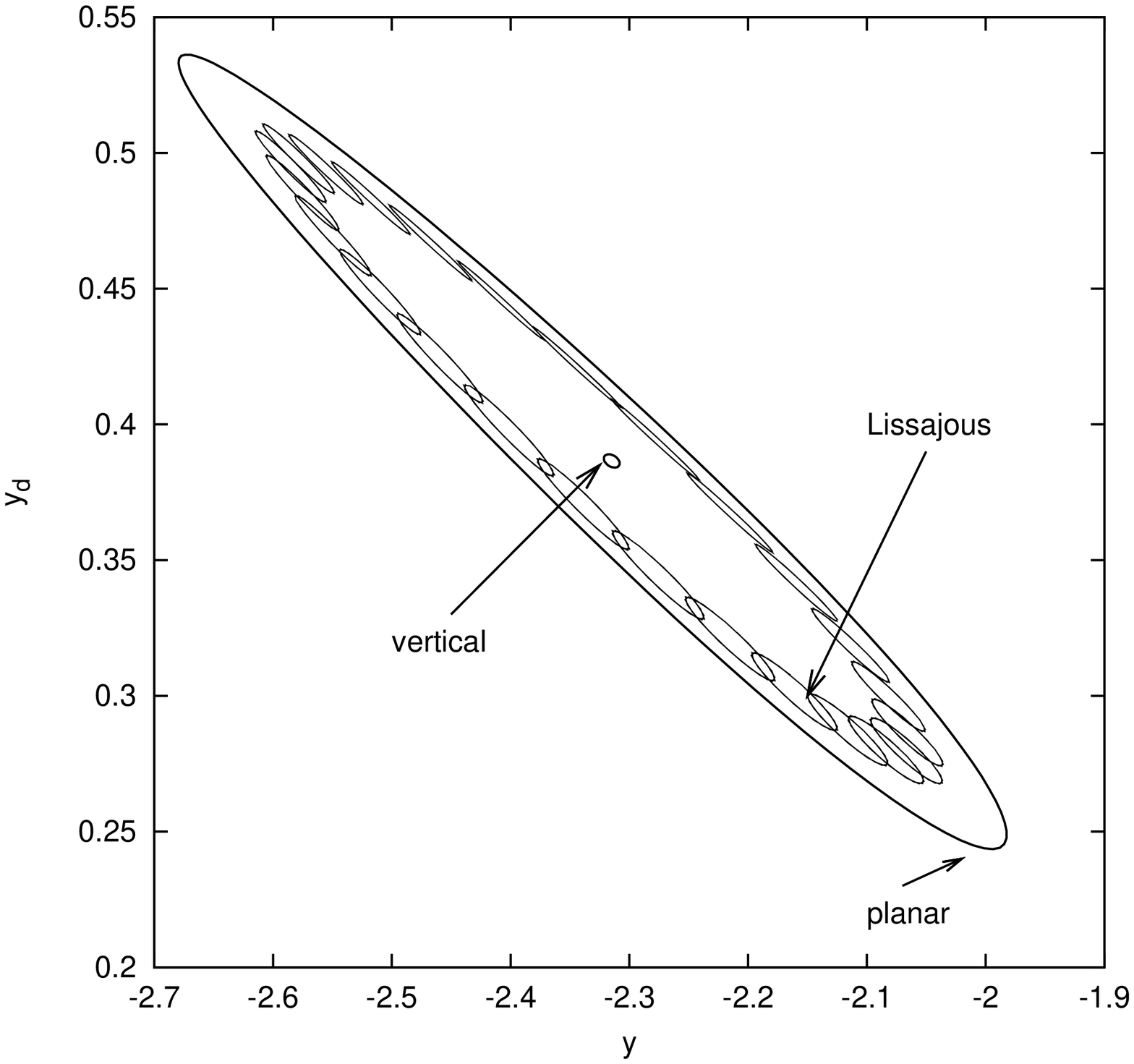}& 
\hspace{-2.cm}\includegraphics[width=60mm,angle=-90]{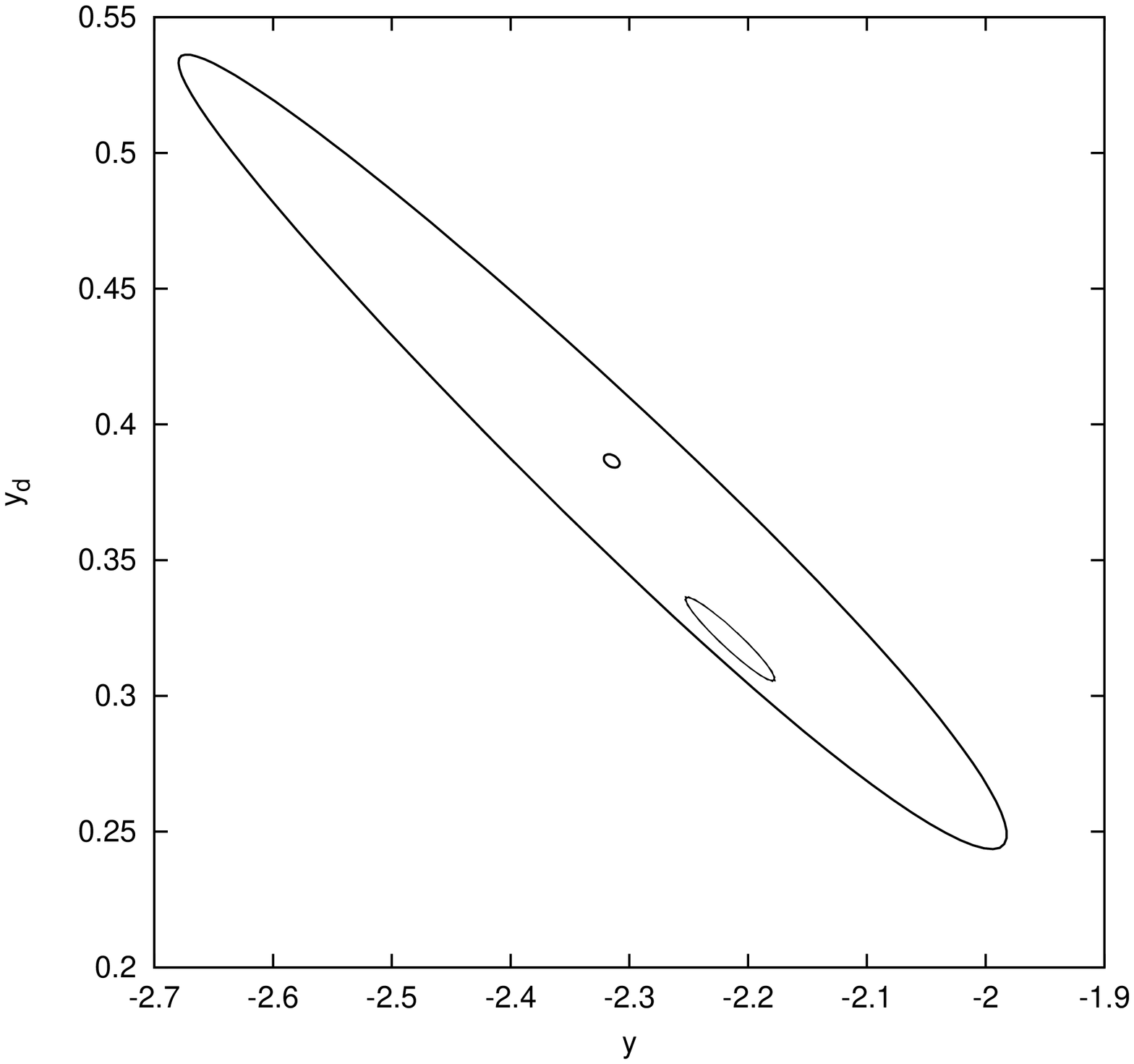}\\
\hspace{-1.cm}\includegraphics[width=60mm,angle=-90]{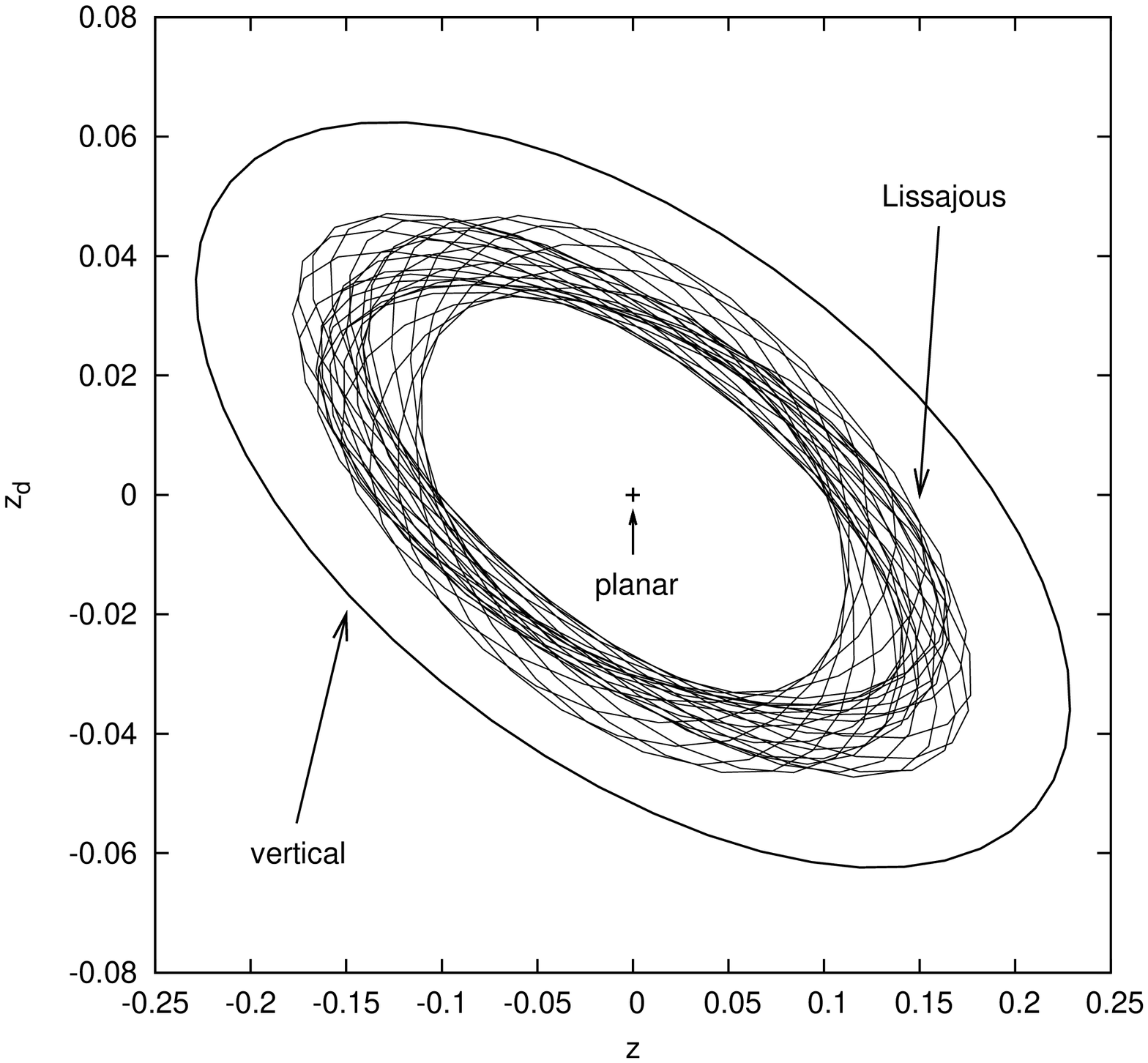}& 
\hspace{-2.cm}\includegraphics[width=60mm,angle=-90]{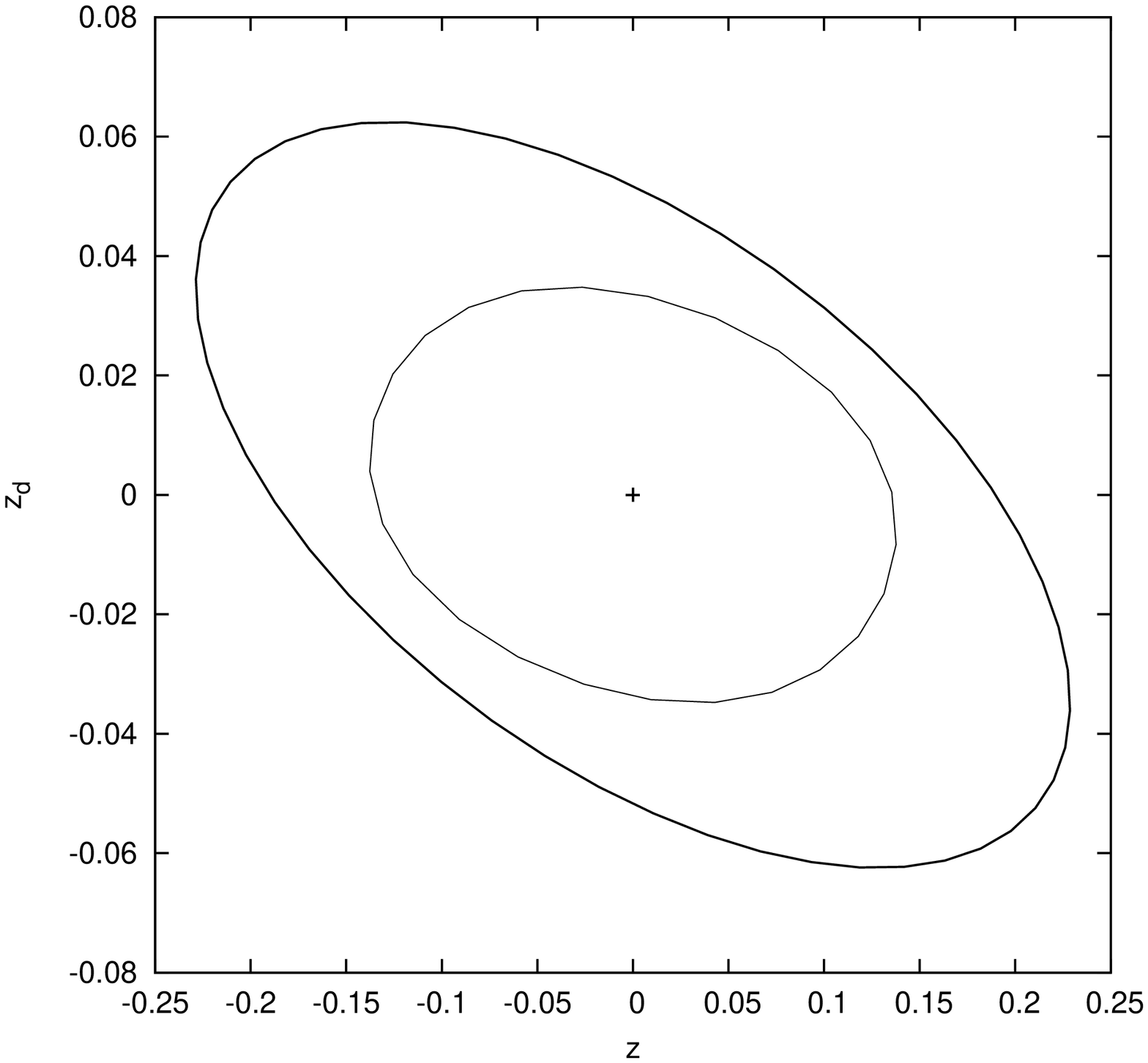} 
\end{tabular}
\caption{\sf $(y,\dot{y})$ and $(z,\dot{z})$ projections of the intersection
with the $\{x=0\}$ plane of the stable manifolds associated with different types of
orbits of energy $C=3.163$, around the $L_2$ point. On the left, on the top one
can see that the projection corresponding to the manifold associated with the
planar Lyapunov orbit is a single closed curve which contains the projections of all the other central
orbits; on the bottom, the projection corresponding to the
manifold associated with the vertical Lyapunov orbit is again a single closed curve which contains all the other
projections. On the right, we represent the same behaviour, underlining that each closed curve constituting the projection of the hyperbolic manifold associated with the Lissajous orbit corresponds to a fixed value of one of the two phases.}
\label{fig:2}
\end{center}
\end{figure}

The stable and unstable manifolds of the periodic orbits are 2--dimensional. Once a
displacement $\epsilon$ has been selected, given a point $\mathbf{x}_{0}(0)$ on
the periodic orbit, $\mathbf{x}_{S,U}(t)$, $t\in [0,T]$, provide initial
conditions on the stable/unstable manifolds. In this way, $\mathbf{x}_{0}(t)$,
$t\in [0,T]$, can be thought as one of the parameters that generate the
manifolds. It is usually called the \emph{parameter along the orbit} or
\emph{phase}. Sometimes it will be convenient to normalise the period so that $t\in
[0,2 \pi]$.  The other parameter is the elapsed time for going, following the
flow with increasing/decreasing $t$, from the initial condition to a certain
point on the manifold. This time interval is usually called the \emph{parameter
along the flow}.

We remark that this parametrization depends on the choice of $\epsilon$ and on
the way in which the stable/unstable direction is normalized. A small change in
$\epsilon$ produces an effect equivalent to a small change in
$\mathbf{x}_{0}(0)$, in the sense that with both changes we get the same orbits
of the manifold.  Only a small shift in the parameter along the flow will be
observed. This is because the stable/unstable directions are transversal to the
flow.

\subsubsection{Geometric behaviour}\label{sec:2.1.2}

\begin{figure}[b!]
\begin{center}
\begin{tabular}{cc}
\hspace{-1.1cm}\includegraphics[width=60mm,angle=-90]{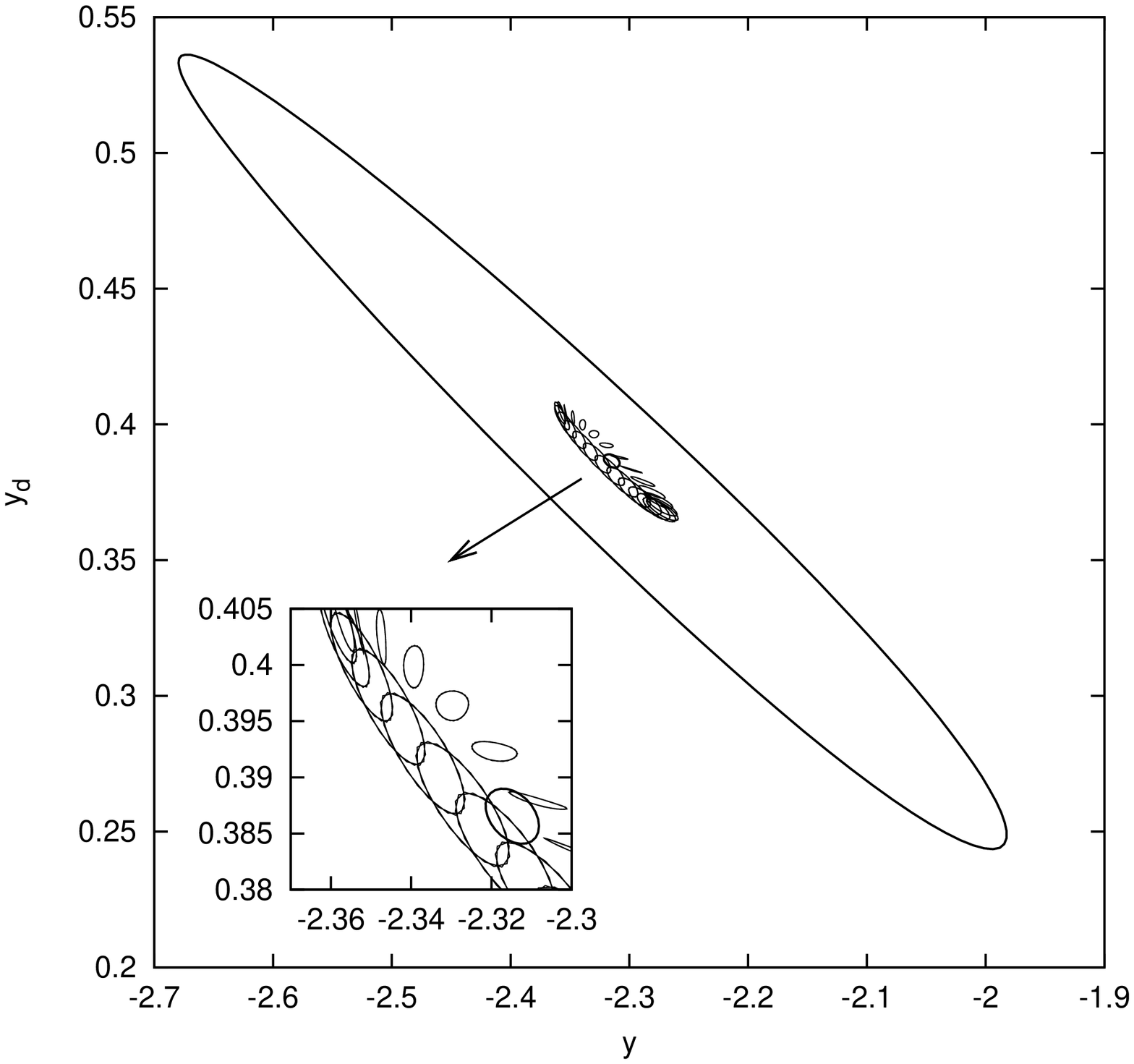}& 
\hspace{-2.0cm}\includegraphics[width=60mm,angle=-90]{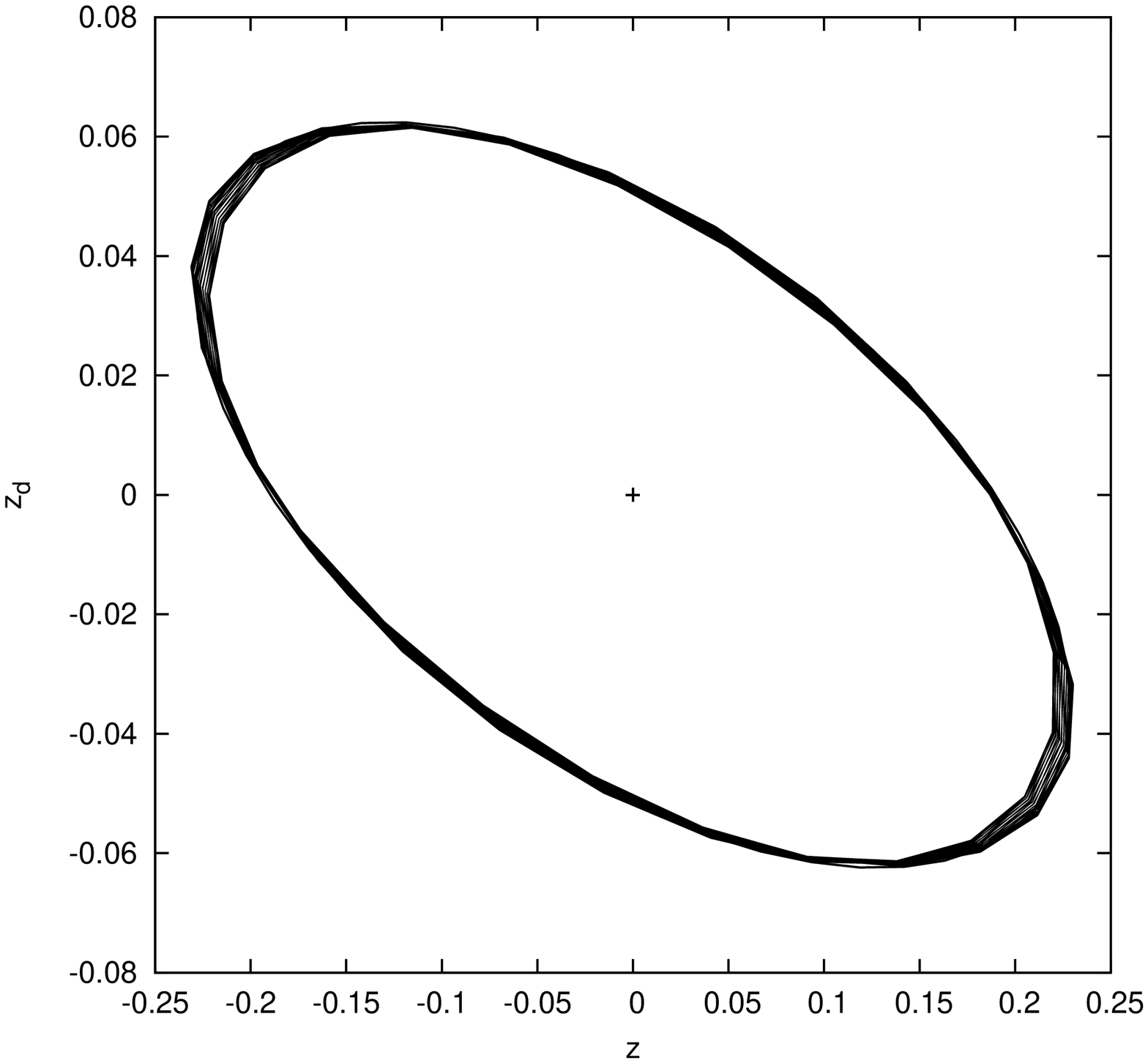}\\
\hspace{-1.1cm}\includegraphics[width=60mm,angle=-90]{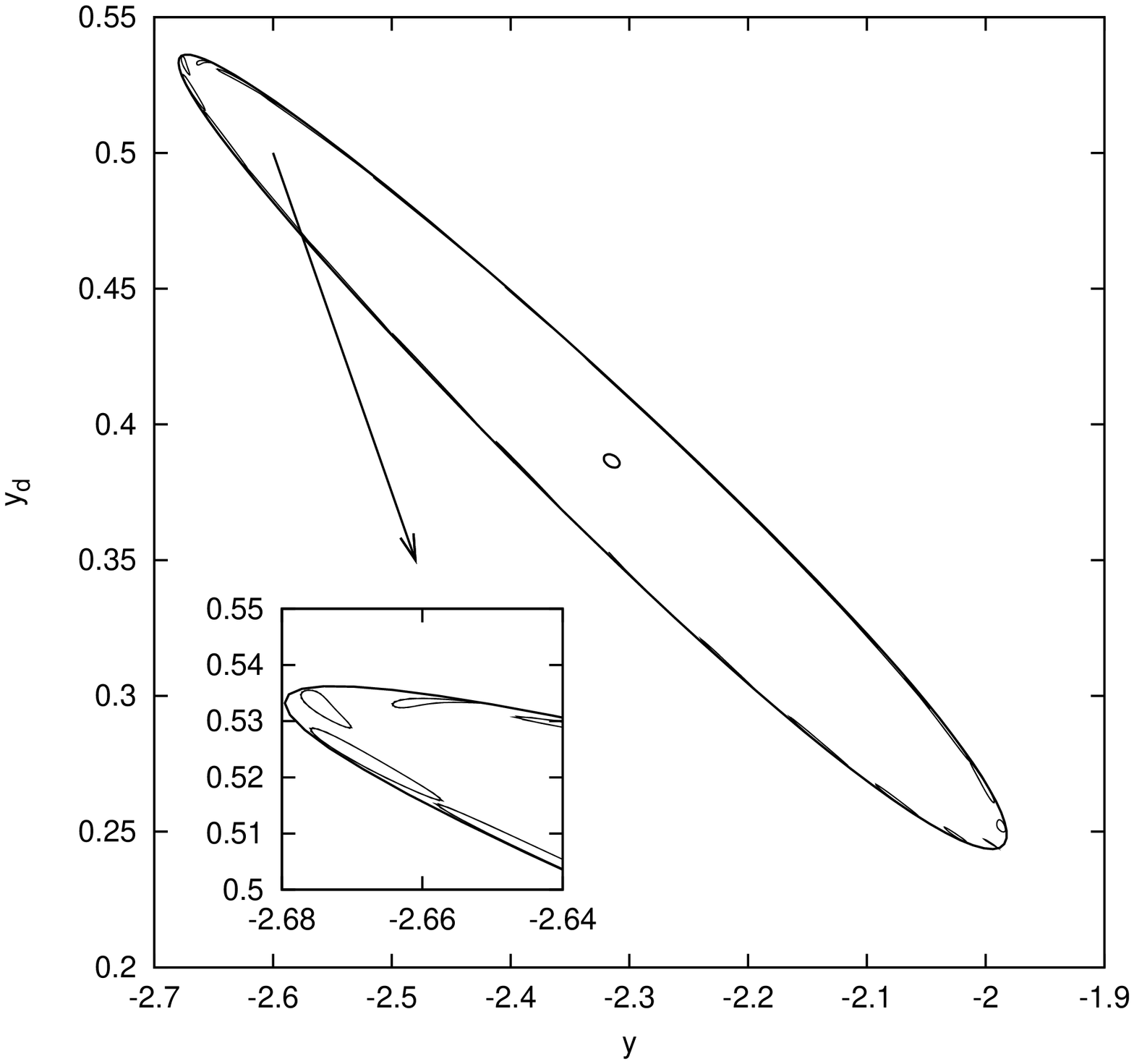}& 
\hspace{-2.0cm}\includegraphics[width=60mm,angle=-90]{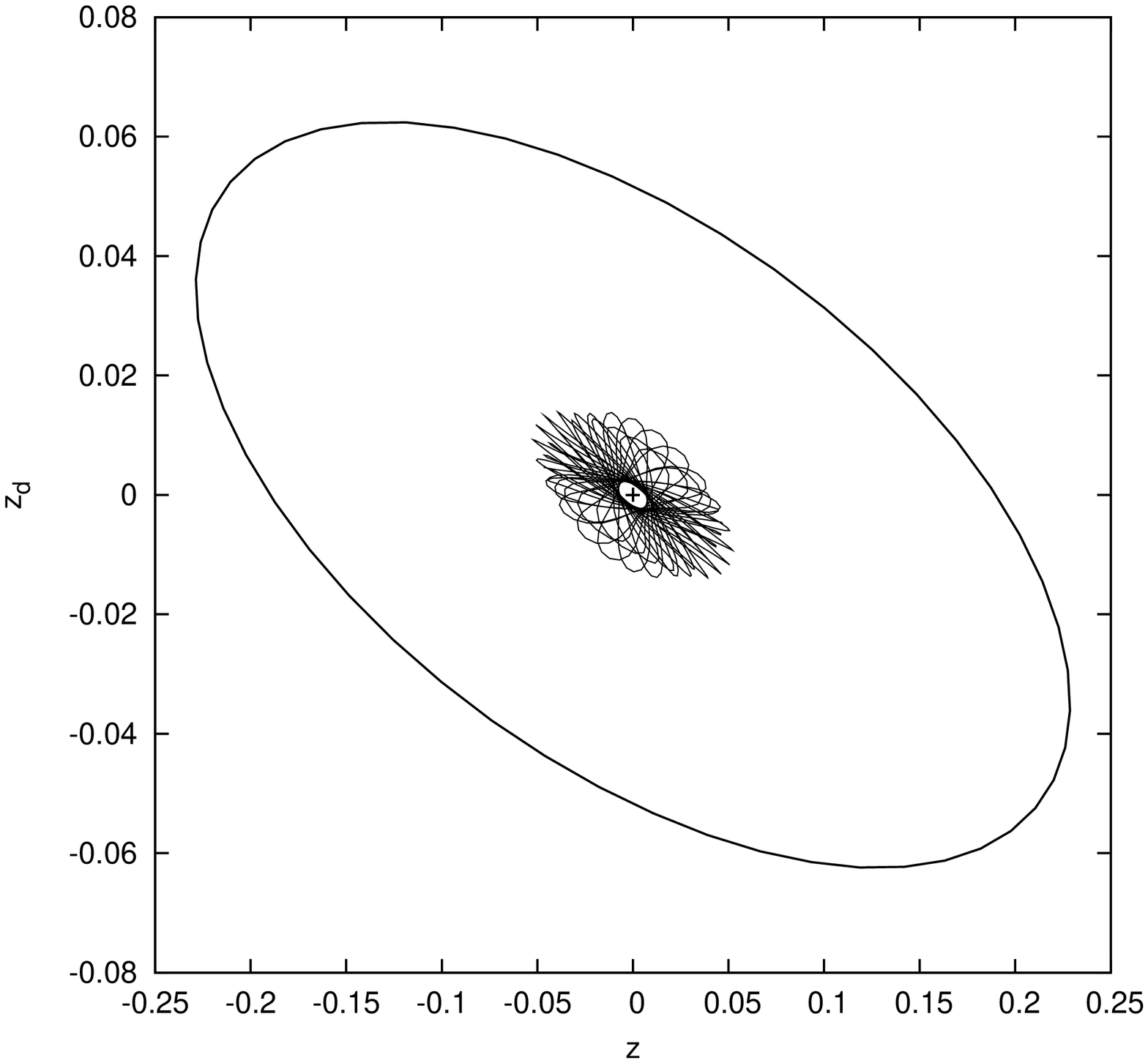} 
\end{tabular}
\caption{\sf $(y,\dot{y})$ and $(z,\dot{z})$ projections of the intersection
with the $\{x=0\}$ plane of the stable manifolds associated with different types of
orbits of energy $C=3.163$, around the $L_2$ point. On the top, the Lissajous orbit has a considerable out--of--plane amplitude (and a very small in--plane one) and thus the projections of the associated stable manifold intersects the ones associated with the vertical Lyapunov periodic orbit. On the bottom, the Lissajous orbit has a considerable in--plane amplitude (and a very small out--of--plane one) and thus the projections of the associated stable manifold intersects the ones associated with the planar Lyapunov periodic orbit.}
\label{fig:3}
\end{center}
\end{figure}

For a fixed value of $C$, around a given equilibrium point there are one planar and one vertical Lyapunov periodic orbit plus several Lissajous orbits of different amplitudes and other types of periodic and quasi-periodic motion, whose existence depends on the energy level considered \cite{GM}. 

Let us consider, for a well-defined energy level, the intersection between the
stable invariant manifold associated with different central orbits and the
$\{x=0\}$ plane (see Fig. ~\ref{fig:2}). In the $(y,\dot y)$ and
$(z,\dot z)$ projections, the hyperbolic manifold associated with the vertical
Lyapunov periodic orbit gives rise to a single closed curve, the one associated
with the planar Lyapunov periodic orbit generates, respectively, a single
closed curve and a point at the origin. On the other hand, in both projections
the hyperbolic manifold associated with a Lissajous orbit, which has dimension 3, produces an annular
region, composed by infinitely many closed curves chained together. Clearly, this is
because a periodic orbit is a $\mathbb{S}^1$ object, a Lissajous orbit is a
$\mathbb{T}^2$ one. If we fix the value of one of the two phases, say $\phi_1$,
characterizing a Lissajous orbit, and let the value of the other, say $\phi_2$,
to vary in $[0,2\pi]$ we get one of the closed curves forming the annular
region, as shown in Fig.~\ref{fig:2} on the right. 

Keeping constant the value of $C$, distinct Lissajous orbits are found by increasing the out-of-plane amplitude and decreasing the in-plane one or viceversa. The $(y, \dot y)$ and $(z,\dot z)$ projections corresponding to the hyperbolic manifolds associated with different Lissajous orbits with similar values of the amplitudes may intersect each other, but they tend to stay one inside the other. Furthermore, the projections of the hyperbolic manifold associated with the Lissajous orbit with greater out-of-plane amplitude will be closer to the projections associated with the vertical Lyapunov periodic orbit. In the limit case, when the Lissajous orbit takes vertical amplitude almost as big as that of the vertical Lyapunov periodic orbit, the two projections overlap. The same argument holds with respect to the planar Lyapunov periodic orbit when increasing the in-plane amplitude. This is shown in Fig.~\ref{fig:3}. 

If we consider other sections or other types of central orbits apart from the Lissajous ones, the same qualitative behaviour is found. In turn, the role of outer bound is played by the hyperbolic manifold associated with the planar Lyapunov periodic orbit in the $(y,\dot y)$ projection, by the one
associated with the vertical Lyapunov periodic orbit in the $(z, \dot z)$
plane. This result is analogous to the well-known Poincar\'e map representation
of the central manifold dynamics (see Fig.~\ref{fig:2_new}). This explains how the hyperbolic manifolds associated with planar and vertical Lyapunov orbits act as energy boundaries for transit orbits lying inside $\mathcal{W}^{s/u}(L_i)$.

\subsubsection{Initial conditions corresponding to transit orbits in $\mathcal{W}^s(L_2)$}\label{sec:2.1.3}

Following the above considerations, we carry on the
computation of transit trajectories belonging to  $\mathcal{W}^s(L_2)$ as follows. First, for a fixed value of the Jacobi constant, we compute the planar and the
vertical Lyapunov periodic orbit around the libration point, as well as the initial conditions determining the proper branch (the one that escapes from the Earth -- Moon neighbourhood backwards in time) of their
corresponding stable manifold. The initial conditions on the
hyperbolic manifolds of both orbits are integrated backwards in time,
up to their first intersection with the $\{x=0\}$ section. This procedure yields two closed curves, as already shown
in Figs.~\ref{fig:2} and \ref{fig:3}.

Next, instead of setting a grid within each closed curve, we look for an uniform distribution of random initial
conditions lying inside them. By means of a Knuth shuffle algorithm \cite{K} (see Appendix) we generate a pair of random numbers for
$(y,\dot y)$; if the point is inside the closed curve in the $(y,\dot y)$ plane,
then we generate another pair of random numbers, $(z,\dot z)$, that now should
stay in the interior of the closed curve of the $(z,\dot z)$ projection. When both pairs
fulfill the requirements, we complete the initial conditions setting $x=0$  and
determining the $\dot x$ coordinate by means of the Jacobi first integral.

We notice that taking initial conditions on the $\{x=0\}$ section means assuming the asteroids to have already left the Main Asteroid Belt and to have started moving in the Earth -- Moon neighbourhood. Other sections apart from the $\{x=0\}$ one could have been selected. In the future, we will use the same methodology with a different choice, depending on the aspect we want to emphasize. For instance, the main source of colliding orbits to belong to the ecliptic plane.

\section{Impacts coming from $\mathcal{W}^s(L_2)$}\label{sec:3}

The first exploration performed consists in numerically integrating the equations of motion of the CR3BP starting from the initial conditions derived as just explained. We take $10^5$ points inside the $(y,\dot y)$ curve and, for each
of these points, 10 pairs of $(z,\dot z)$ coordinates. Hence, for each energy level, we explore the behaviour of $10^6$ trajectories. This selection results in a distribution of initial conditions corresponding to transit orbits, which is uniform both in $(y,\dot y)$ and $(z,\dot z)$ coordinates. However, a different matching among coordinates would be possible.
\begin{figure}[t!]
\begin{center}
\begin{tabular}{cc}
\hspace{-1.5cm}\includegraphics[width=75mm]{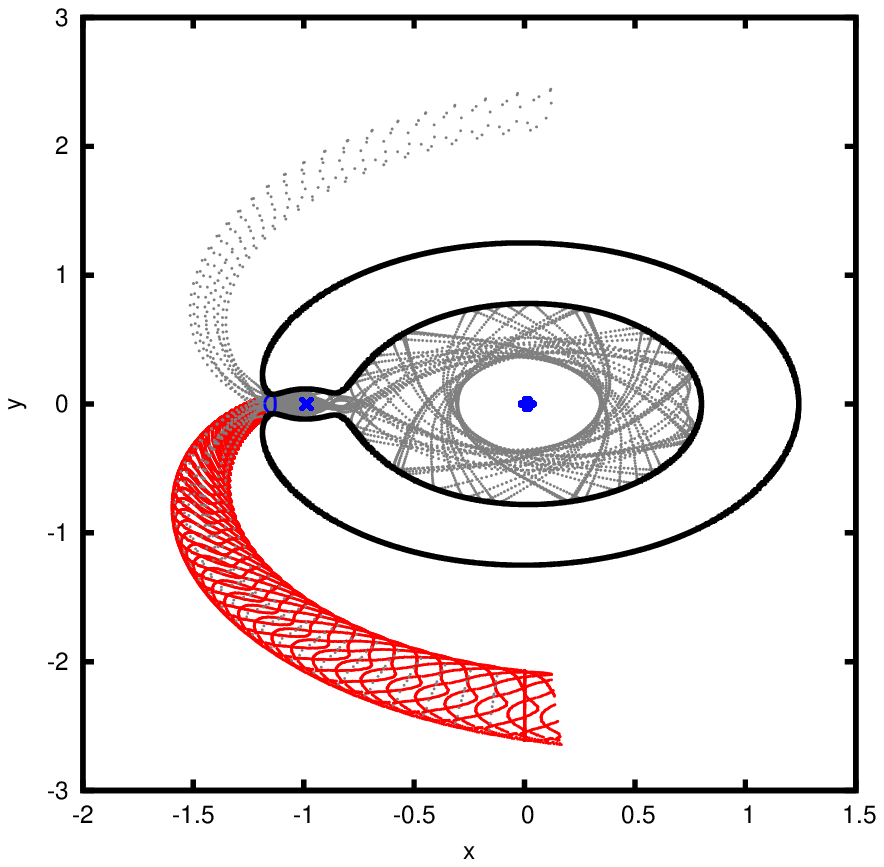}& 
\hspace{-1.0cm}\includegraphics[width=75mm]{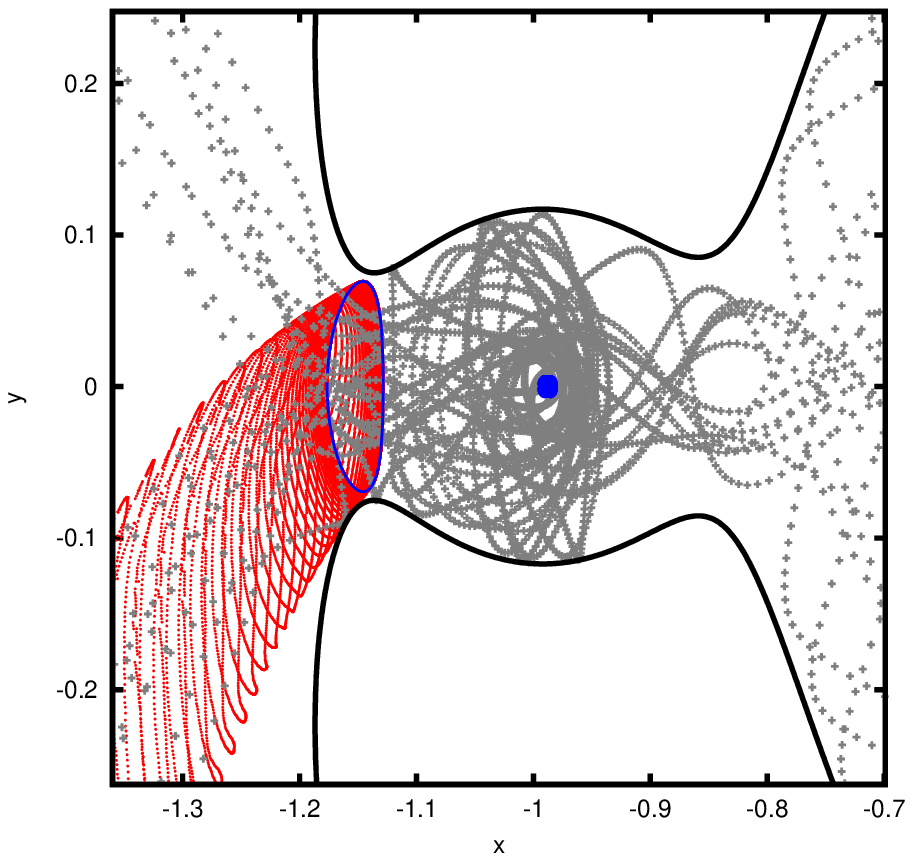}
\end{tabular}
\caption{\sf Schematic representation of the methodology implemented. The red tube is $\mathcal{W}^s(Planar~Lyapunov)$, the gray trajectories inside it are transit orbits, in blue we display the planar Lyapunov periodic orbit, the Moon and the Earth. The trajectories are permitted to live out of the area bounded by the Hill's region. On the right, we show a closer view of the dynamics around the Moon.}
\label{fig:4}
\end{center}
\end{figure}

Once again, we stress that the trajectories corresponding to such initial conditions will be driven by the stable component of $\mathcal{W}^c(L_{2})$, without lying on $\mathcal{W}^s(L_{2})$. They stay inside the dynamical tube generated by $\mathcal{W}^s(L_{2})$ \cite{GKLMMR}.  

The procedure is implemented for 10 equally spaced energy levels $C$ in the range $C_3 < C < C_2$. In Fig.~\ref{fig:4}, we show the Hill's region this energy range corresponds to, the boundary of $\mathcal{W}^s(L_2)$ in red and some transit trajectories in gray.

As the more intense lunar bombardment took place some billions years ago, we must consider different values for the Earth -- Moon distance $d_{EM}$. Indeed, the Moon is receding from the Earth: the rate of recession has not been constant in the past and it did not behave linearly either (see, for example, \cite{G,TMV,MA,LF}). We take $4$ values for $d_{EM}$, 232400, 270400, 308400, 384400 km, respectively. According to \cite{LF}, they correspond approximately to 4., 3.4, 2.5 and 0 Gy ago.

The maximum allowed time for impacting onto the surface of the Moon is 60 years, provided the assumption of a no longer life in the region under consideration.
Within this time span, we get a numerical evidence that the minor bodies can behave in one of the following ways:
\begin{enumerate}
\item[(1)] they collide with the Moon without overcoming the $L_1$ border;
\item[(2)] they collide with the Moon after overcoming the $L_1$ border and thus performing several loops around the Earth;
\item[(3)] they keep wandering around the Earth inside the area delimited by the zero-velocity surface;
\item[(4)] they escape from the Earth -- Moon neighbourhood just after jumping on the $L_2$ gate;
\item[(5)] they exit from the Earth -- Moon neighbourhood after wandering for a certain interval of time around the Earth.
\end{enumerate}
Note that just the first two cases cause the formation of craters of impact on the surface of the Moon.

\begin{figure}[t!]
\begin{center}
\begin{tabular}{cc}
\hspace{-1.5cm}\includegraphics[width=53mm,angle=-90]{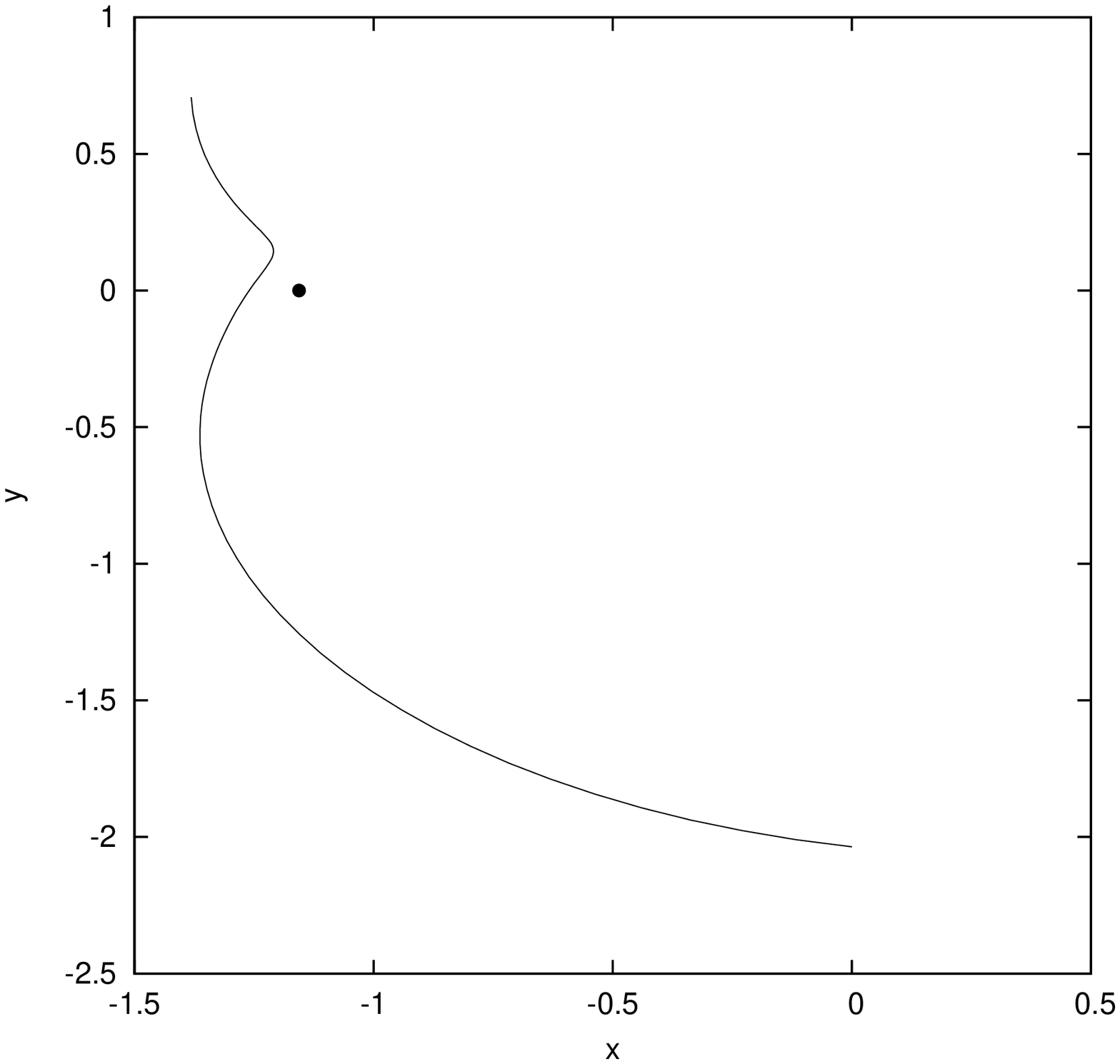}& 
\hspace{-1.0cm}\includegraphics[width=53mm,angle=-90]{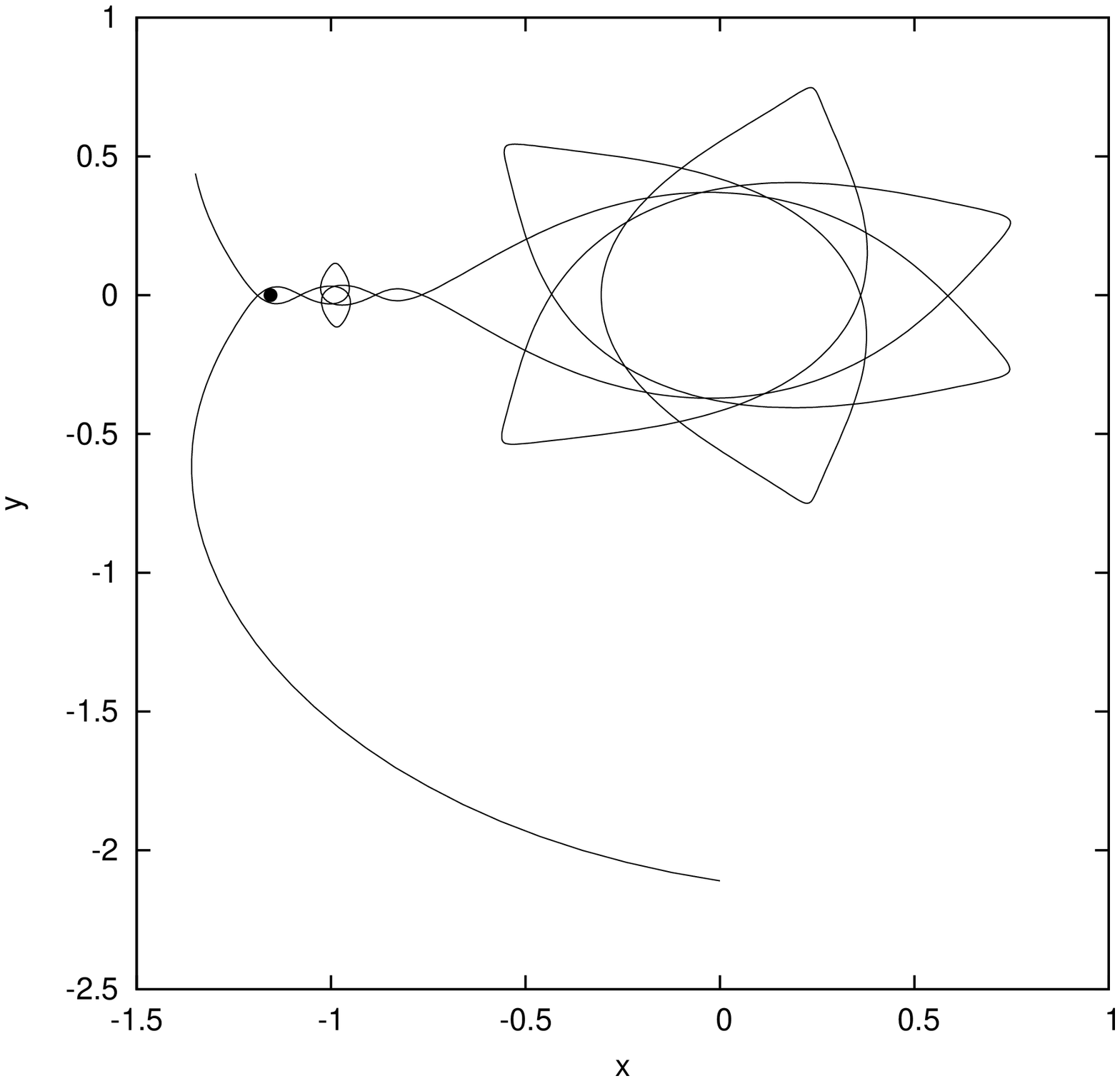}
\end{tabular}
\caption{\sf We display the two ways in which a particle can escape from the Earth -- Moon neighbourhood. On the left, the asteroid jumps on the $L_2$ gate; on the right, it performs some loops around the Earth and then joins $\mathcal{W}^u(L_2)$ to escape.}
\label{fig:5}
\end{center}
\end{figure}

If a trajectory collides with the Moon we calculate the longitude and latitude corresponding to the site of impact, the velocity, the angle and the orbital elements of the osculating ellipses with respect to the Earth at the initial condition. We recall that we neglect the Moon's orbital inclination with respect to the Earth's orbit.

Regarding the fourth and the fifth item (see Fig.~\ref{fig:5}), the mechanism of escaping is produced by $\mathcal{W}^u(L_2)$. In the future, it will be interesting to analyse how the impact phenomena are fostered by homoclinic connections associated with $L_2$. By finding succeeding intersections, in a given energy level, between the stable and the unstable manifold associated with the planar Lyapunov periodic orbit and simultaneously between the stable and the unstable manifold associated with the vertical Lyapunov one, it is possible to construct cycling paths, which bring the particle in and out the region demarcated by the zero-velocity curve. However, preliminary simulations suggest the percentage of impacts offered by this loop to be small in comparison with the total amount of collisions.

\subsection{Results}\label{sec:3.1}

We can highlight the following outcomes:
\begin{itemize} 
\item[(a)] the percentage of impacting orbits over all the initial conditions launched is 13$\%$;
\item[(b)] the smaller $d_{EM}$, the greater the above percentage;
\item[(c)] the amount of particles that still wander around the Earth inside the zone bounded by the zero-velocity surface after 60 years is 0.1$\%$;
\item[(d)] in all the cases of collision, the heaviest probability of impact takes place at the apex of the lunar surface $(90^{\circ} W, 0^{\circ})$.
\end{itemize}
Point (a) and (c), in particular, reveal that the 60 years assumed are not restrictive with respect to a lunar impact. We also notice that in the time interval considered the most of the asteroids escapes from the region we are interested in. As already mentioned, it looks like just few of them are able to go back to the Earth -- Moon neighbourhood later. It is reasonable to think that they remain in the Inner Solar System and occasionally are pushed towards the Earth again. 

\begin{figure}[h!]
\begin{center}
\begin{tabular}{cc}
\includegraphics[width=80mm]{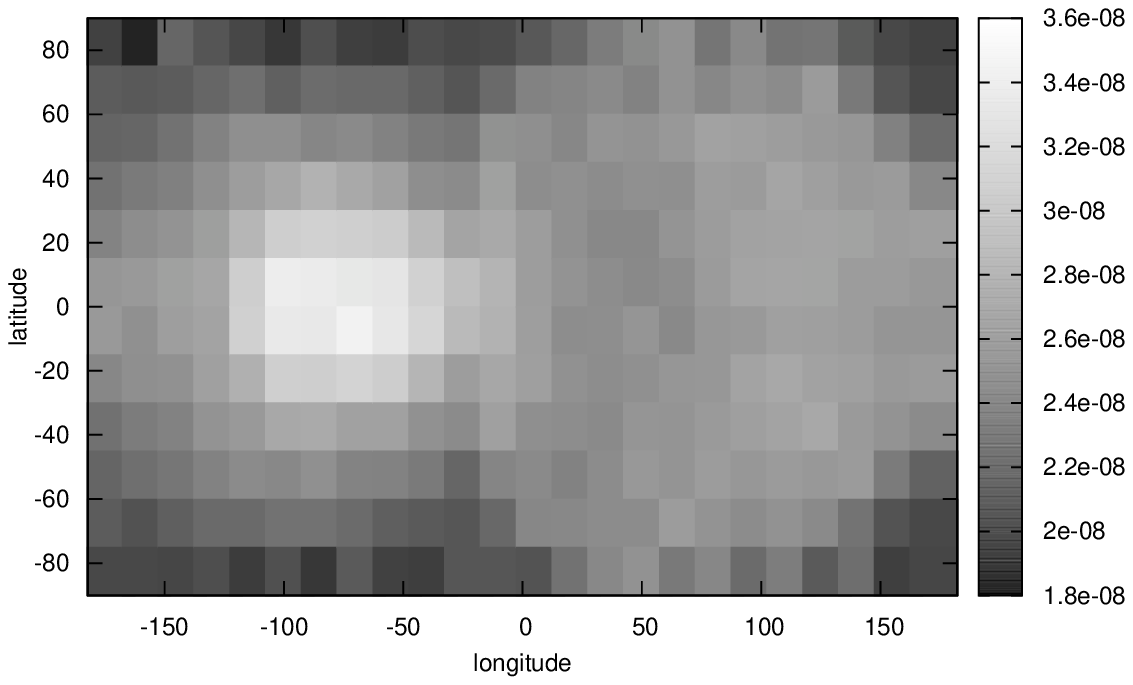}& 
\includegraphics[width=80mm]{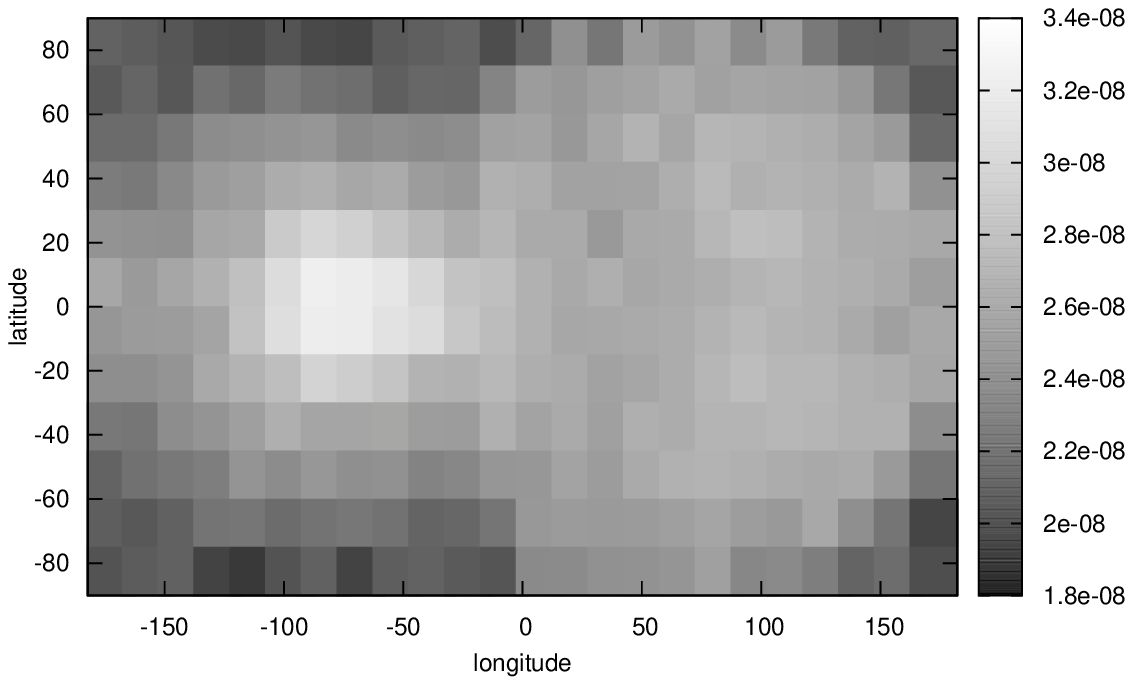}\\
\includegraphics[width=80mm]{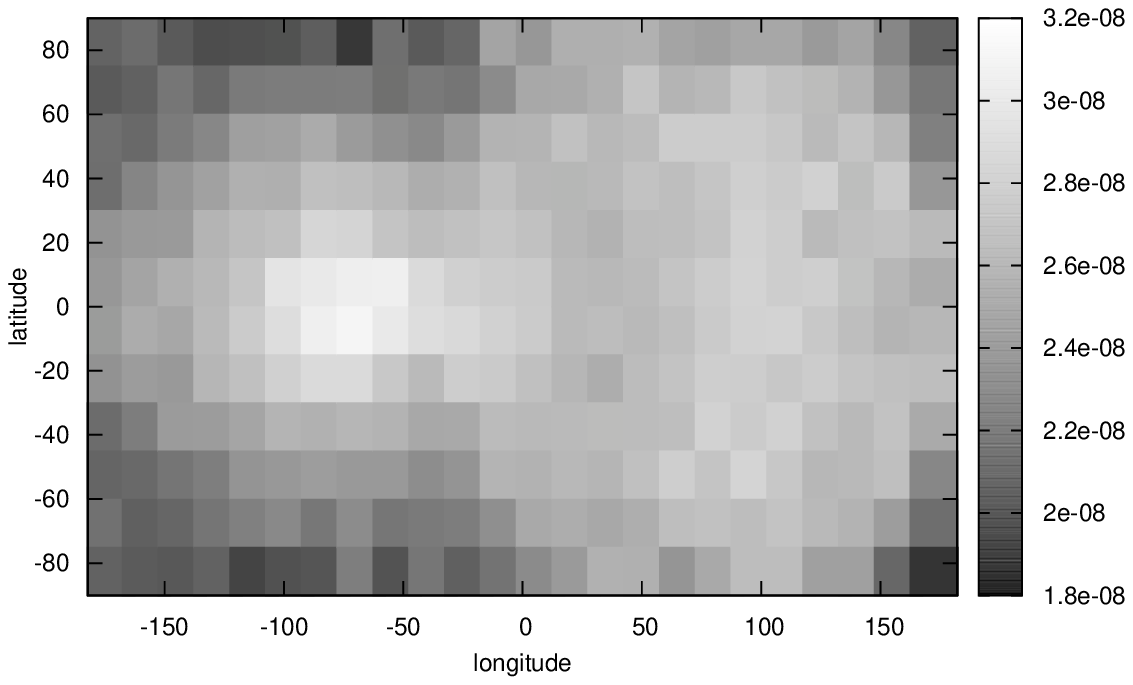}& 
\includegraphics[width=80mm]{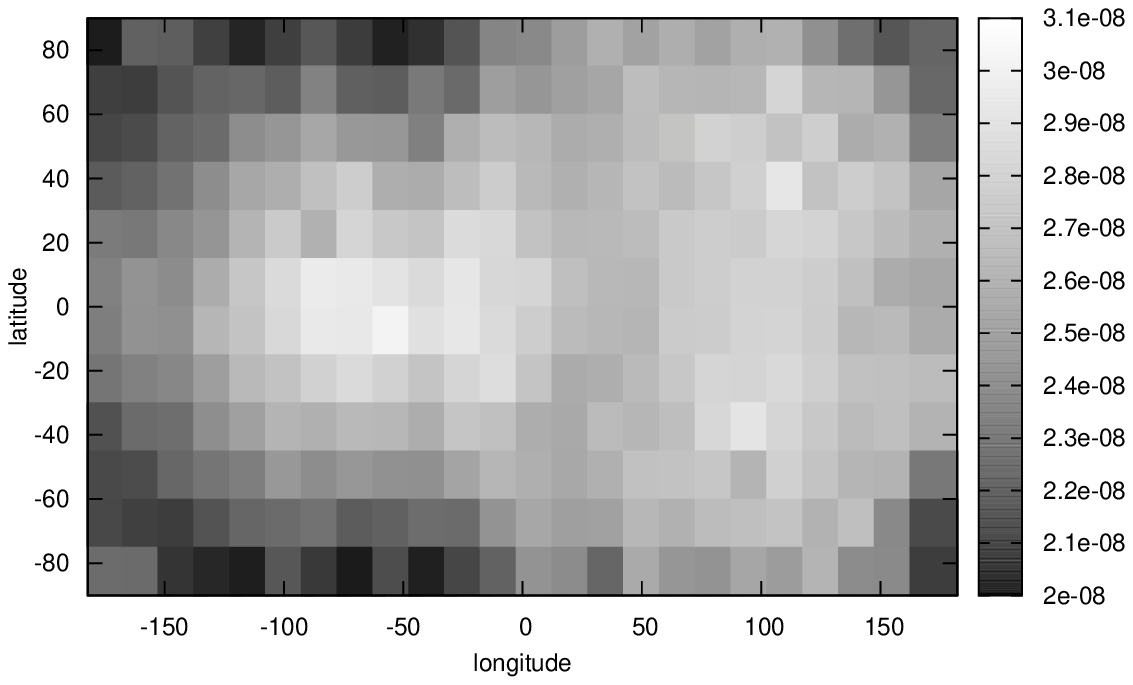} 
\end{tabular}
\caption{\sf We display the density of impact (number of impacts per unit of area normalized with respect to the total number of impacts obtained) computed by exploiting the CR3BP equations of motion. The surface of the Moon is discretized in squares of $15^{\circ}\times 15^{\circ}$ and 4 different values for the Earth -- Moon distance are considered.  On the top, $d_{EM}=232400$ km and $d_{EM}=270400$ km; on the bottom, $d_{EM}=308400$ km and $d_{EM}=384400$ km. The color bar indicates that the lighter the color of the square the greater the impact density.}
\label{fig:6}
\end{center}
\end{figure}

\begin{figure}[h!]
\begin{center}
\begin{tabular}{cc}
\includegraphics[width=80mm]{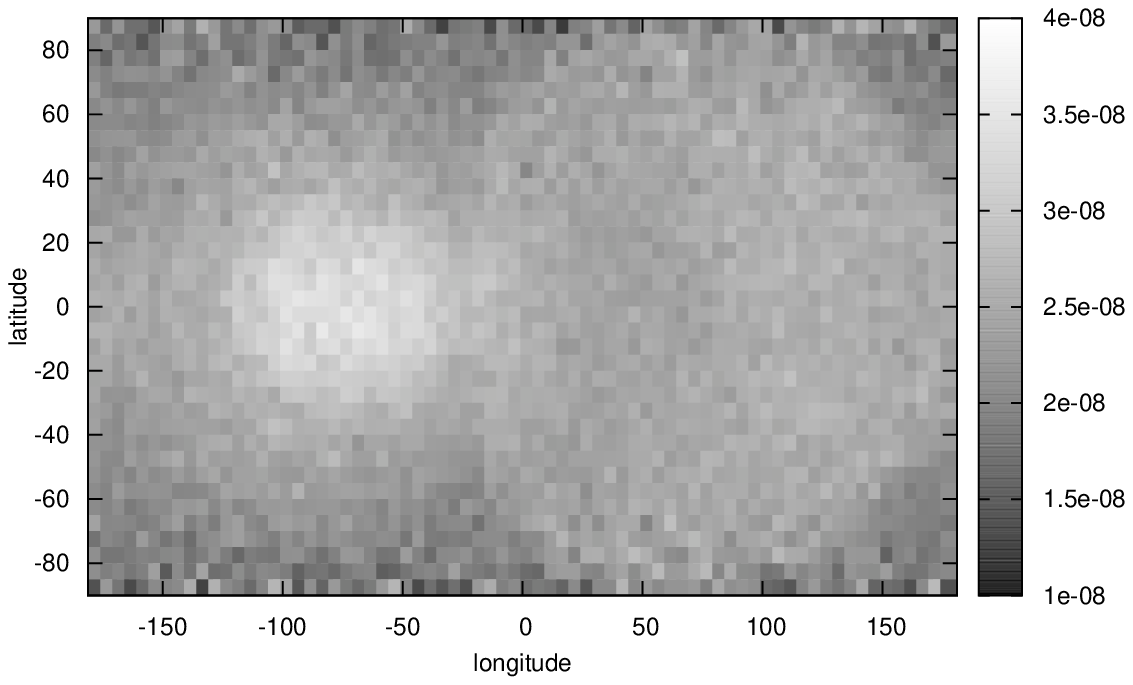}& 
\includegraphics[width=80mm]{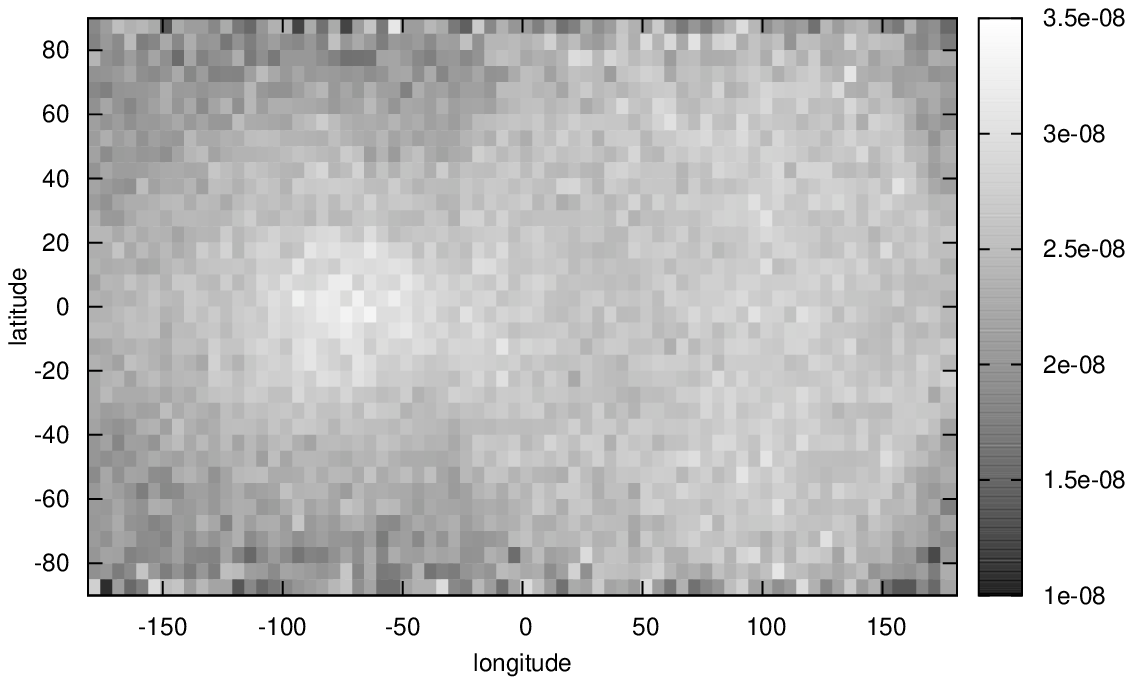}
\end{tabular}
\caption{\sf We display the density of impact (number of impacts per unit of area normalized with respect to the total number of impacts obtained) computed by exploiting the CR3BP equations of motion. The surface of the Moon is discretized in squares of $5^{\circ}\times 5^{\circ}$. On the left, $d_{EM}=232400$ km; on the right, $d_{EM}=308400$ km. The color bar indicates that the lighter the color of the square the greater the impact density.}
\label{fig:7_new_a}
\end{center}
\end{figure}

\begin{figure}[h!]
\begin{center}
\begin{tabular}{cc}
\includegraphics[width=80mm]{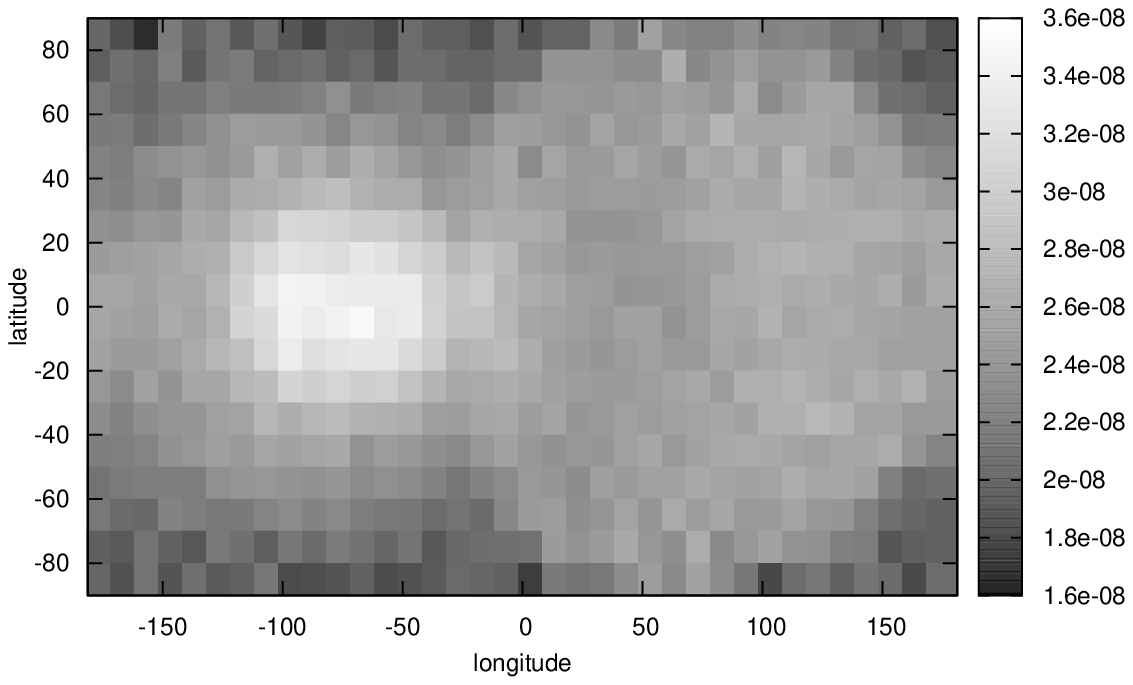}& 
\includegraphics[width=80mm]{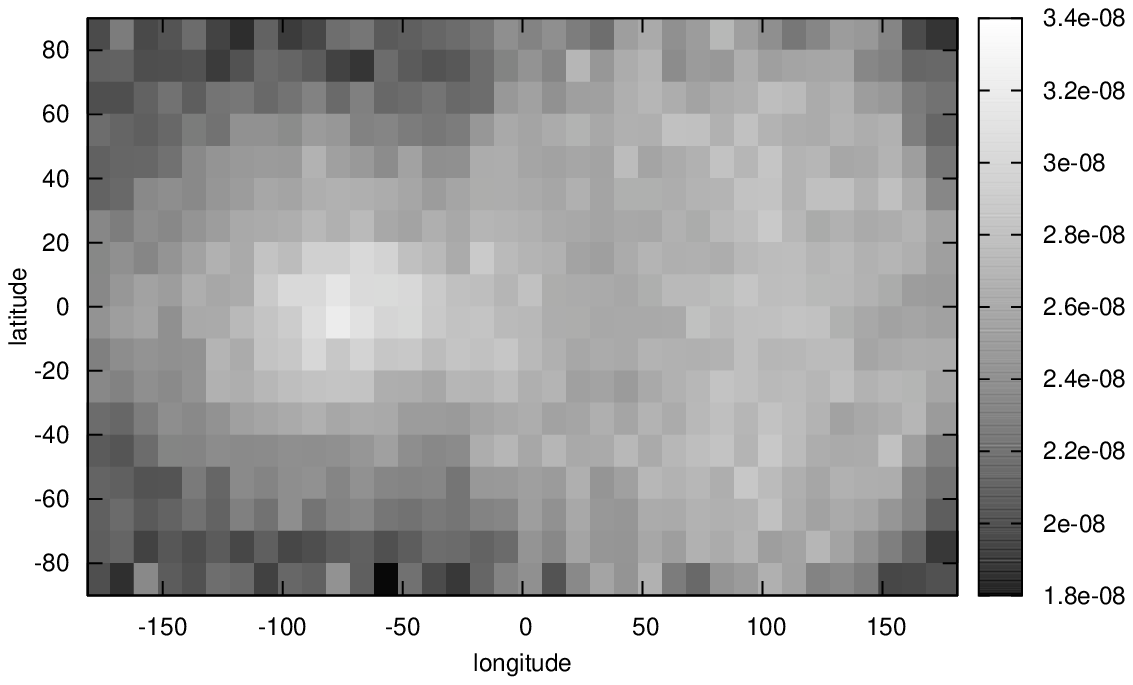} 
\end{tabular}
\caption{\sf We display the density of impact (number of impacts per unit of area normalized with respect to the total number of impacts obtained) computed by exploiting the CR3BP equations of motion. The surface of the Moon is discretized in squares of $10^{\circ}\times 10^{\circ}$. On the left, $d_{EM}=232400$ km; on the right, $d_{EM}=308400$ km. The color bar indicates that the lighter the color of the square the greater the impact density.}
\label{fig:7_new_b}
\end{center}
\end{figure}

\begin{figure}[bhtp!]
\begin{center}
\begin{tabular}{cc}
\includegraphics[width=80mm]{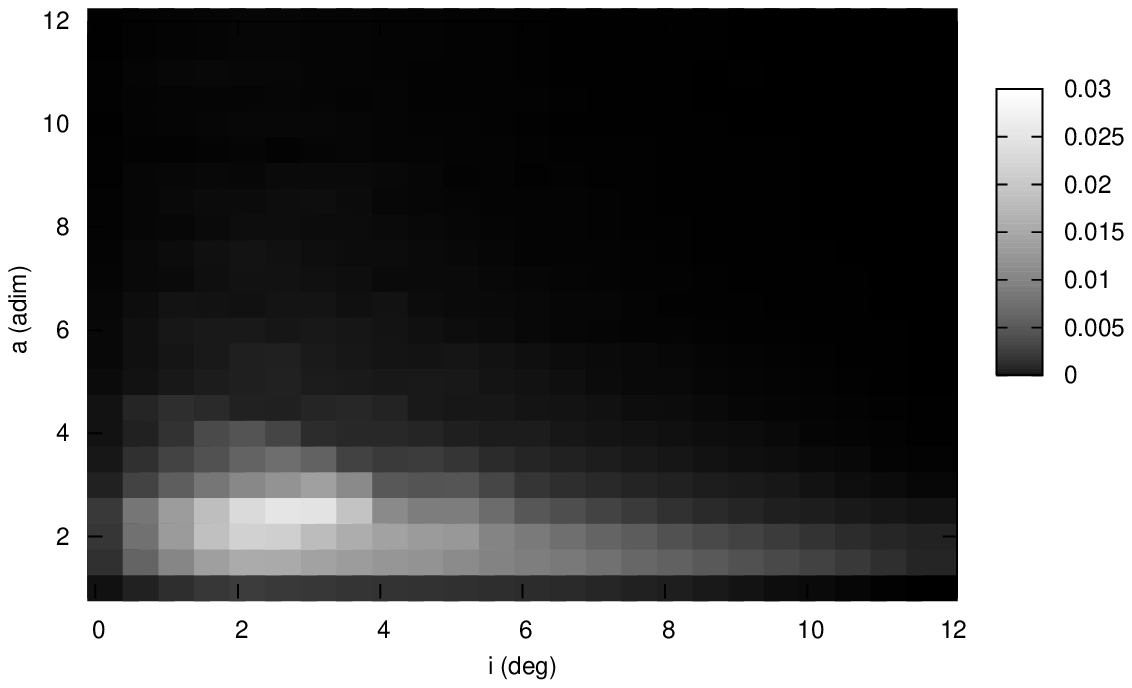}& 
\includegraphics[width=80mm]{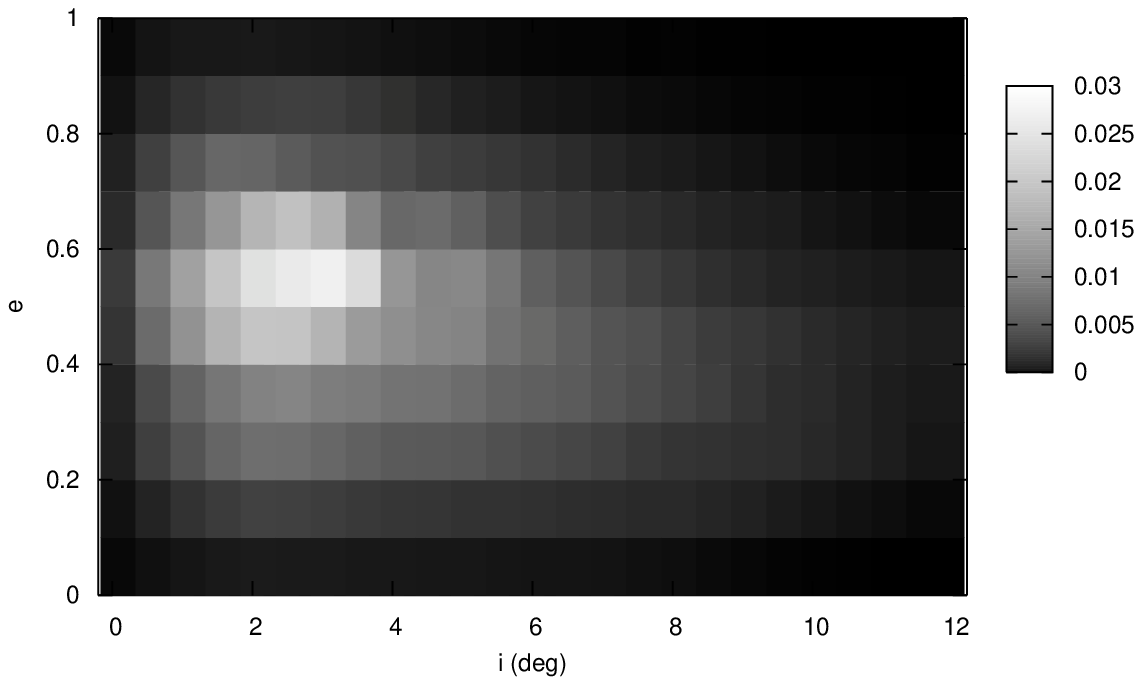}
\end{tabular}
\caption{\sf For each initial condition belonging to $\mathcal{W}^s(L_2)$ and colliding with the lunar surface, we compute the orbital elements which correspond to the osculating ellipse at $t=0$ with respect to the Earth. The plot on the left shows the probability of impact considering as variables the inclination $i$ and the semi-major axis $a$; the one on the right takes as variables $i$ and the eccentricity $e$. The color bar refers to the number of impacts normalized with respect to the total number of impacts found. The lighter the color the greater the probability. For these plots the Earth -- Moon distance is assumed to be $d_{EM}=384400$ km. The $i$ and $a$ ranges are discretized at steps of $0.5$ degrees and $d_{EM}$, respectively. The $e$ range is discretized at steps of $0.1$.}
\label{fig:7}
\end{center}
\end{figure}

In Fig.~\ref{fig:6}, we represent the surface of the Moon in terms of longitude and latitude. In particular, we discretize the lunar sphere in squares of $15^{\circ}\times 15^{\circ}$. We notice different colors associated with each square: the lighter the color the greater the number of collisions per unit of area normalized with respect to the total number of impacts obtained. We remark that to consider another discretization of the lunar surface would not bring any relevant difference from a qualitatively point of view (see Figs.~\ref{fig:7_new_a} and ~\ref{fig:7_new_b}).

As said, we compute the orbital elements of the osculating ellipses at $t=0$ with respect to the Earth, corresponding to initial conditions leading to impact. This means that every set $(x,y,z,\dot x,\dot y,\dot z)$ providing a collision with the Moon is transformed to inertial coordinates centered at the Earth and these are then turned into orbital elements. In particular, we get the semi-major axis $a$, the eccentricity $e$, the inclination $i$, the longitude of the ascending node $\Omega$, the argument of perigee $\omega$ and the true anomaly $\nu$. We notice that the initial conditions are taken far enough (at least about $500000$ km if we assume $d_{EM}=384400$ km) from the Moon to be allowed to assume a Two--Body approximation and perform this analysis.

 It turns out that the impact is more likely if $(a,e,i)$ belong to the intervals showed in Tab.~\ref{tab:1}. In Fig.~\ref{fig:7}, we display such probabilities for the case $d_{EM}=384400$ km. As before, the lighter the color associated with a given $(i,a)/(i,e)$ square, the greater the probability that such orbital elements would correspond to a colliding trajectory. The probability is normalized with respect to the total number of impacts obtained. 

\begin{table}[h!]\label{tab:1}
\begin{center}
\caption{\sf For each initial condition belonging to $\mathcal{W}^s(L_2)$ and colliding with the lunar surface, we compute the orbital elements which correspond to the osculating ellipse at $t=0$ with respect to the Earth. The impact is more likely if the semi-major axis $a$, the eccentricity $e$ and the inclination $i$ lie in the range shown here.}
\begin{tabular}{c c c}
\hline
\hline
{\bf a ~($d_{EM}$)}&{\bf e}&{\bf i}\\
\hline
\hline\\
\hline
\hline
$[1.5:3]$&$[0.4:0.7]$&$[1.5^{\circ}:3.5^{\circ}]$\\
\hline
\hline
\end{tabular}
\end{center}
\end{table}

\section{Uniform density of lunar impacts: possible paths}\label{sec:4}

In the previous section, we have seen that $\mathcal{W}^s(L_2)$ provides a non-uniform density of impact on the surface of the Moon. The question that naturally arises is where minor bodies producing an uniform distribution of low-energy collisions would come from.

Having this purpose in mind, for every energy level considered in Sec.~\ref{sec:3}, we create a set of initial conditions uniformly distributed on the lunar surface, discretized as before in terms of longitude and latitude. In this case, not only the position coordinates have to be well spread out, but also the velocity ones. We apply the CR3BP equations of motions to such initial conditions backwards in time, up to a maximum of $5$ years, detecting how many trajectories arrive from the $\{x=0\}$ section already mentioned.

To be more precise, if $\gamma\in[-\pi,\pi]$ is a random value of latitude, $\psi\in[0,2\pi]$ a random value of longitude and $\rho_{M}=1737.53/d_{EM}$ the adimensional radius of the Moon, then at $t=0$ $(x,y,z)$ are computed as:
\begin{eqnarray*} 
x=\rho_{M}\cos{(\gamma)}\cos{(\psi)}+\mu-1,\hspace{0.5cm} y=\rho_{M}\cos{(\gamma)}\sin{(\psi)},\hspace{0.5cm} z=\rho_{M}\sin{(\gamma)}.
\end{eqnarray*} 
Both $\gamma$ and $\psi$ are generated with the Knuth shuffle algorithm \cite{K} (see Appendix).

Concerning $(\dot x,\dot y,\dot z)$ at $t=0$, we implement three different procedures for each $15^{\circ}\times 15^{\circ}$ square in order to ensure to fulfill the constraint of uniform distribution on a semisphere of velocities. If $g\equiv(g_x,g_y,g_z)=(x-\mu+1,y,z)$ is the normal at $(x,y,z)$ to the surface vector and $w$ and $h$ are random values in $[0,1]$, the three approaches can be sketched as follows.
\begin{enumerate}
\item[(1)] Let $\beta\in [-\pi,\pi]$ and $\lambda\in [0,\pi]$ be random values. Then
\begin{eqnarray*} 
\dot x=g_x\cos{(\lambda)}\cos{(\beta)},\hspace{0.5cm} \dot y=g_y\cos{(\lambda)}\sin{(\beta)},\hspace{0.5cm} \dot z=-g_z\sin{(\lambda)}.
\end{eqnarray*} 
\item[(2)]  Let $\beta$ be a random value belonging to the interval $[-\pi,\pi]$ and $\lambda=\cos^{-1}{(1-2w)}\in [0,\pi]$. Then
\begin{eqnarray*} 
\dot x=g_x\cos{(\lambda)}\cos{(\beta)},\hspace{0.5cm} \dot y=g_y\cos{(\lambda)}\sin{(\beta)},\hspace{0.5cm} \dot z=-g_z\sin{(\lambda)}.
\end{eqnarray*} 
\item [(3)] Let $\gamma$ and $\psi$ as above, $\xi=2w-1$ and $\eta=2h-1$ such that $-1<\xi,\eta<1$ and $\xi^2+\eta^2<1$. Then
\begin{eqnarray*} 
\dot x=(2\xi\sqrt{1-\xi^2-\eta^2})\cos{(\gamma)}\cos{(\psi)},\hspace{0.5cm} \dot y=(2\eta\sqrt{1-\xi^2-\eta^2})\cos{(\gamma)}\sin{(\psi)},\hspace{0.5cm} \dot z=-[1-2(\xi^2+\eta^2)]\sin{(\psi)}.\
\end{eqnarray*} 
\end{enumerate}
In any case, $(\dot x,\dot y,\dot z)$ are normalized in order to obtain the modulus of the velocity as the one satisfying the chosen $C$.

In this way, we simulate the behaviour of $758640$ particles for each value of $C$, which means $7586400$ particles in total. This value has been chosen to be in agreement with the computations described in Sec.~\ref{sec:3} but also to have an impact density of $2\times 10^{-2}$ km$^{-2}$ for each $15^{\circ}\times 15^{\circ}$ square considered on the surface of the Moon.

\subsection{Results}\label{sec:4.1}
\begin{figure}[ht!] 
\begin{center}
\includegraphics[width=70mm]{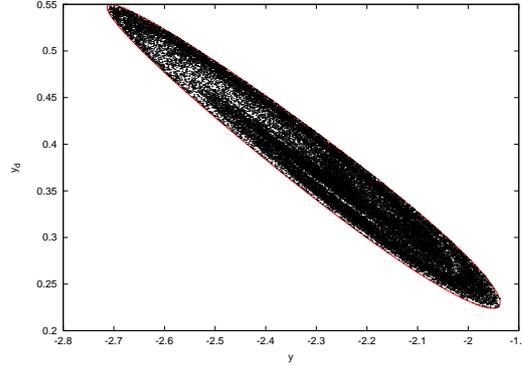}
\caption{\sf If the impact distribution on the surface of the Moon was uniform, initial conditions associated with this pattern inside the $(y,\dot{y})$ curve would collide with the Moon. We remark that the uniform distribution would not be due only to $\mathcal{W}^s(L_2)$, but also to other phenomena. See Figs.~\ref{fig:9} and \ref{fig:10}.}
\label{fig:8}
\end{center}
\end{figure}
The backward simulation reveals that there exist two main dynamical channels leading to lunar collision, for the range of energy under study. In particular, uniform distributed impacts would come either from $\mathcal{W}^s(L_2)$ or from double collision orbits with the surface of the Moon.

In the first case, we notice that all the orbits getting to the $\{x=0\}$ section give rise to points which lie inside the $(y,\dot{y})$ and $(z,\dot{z})$ curves introduced in Sec.~\ref{sec:2.1.2}. This fact can be viewed as a further confirmation of the well-posed procedure adopted previously. With this, we mean that the strategy  defined to determine $\mathcal{W}^s(L_2)$ actually represents the dynamics we were looking for and does not leave out any transit trajectory. The interesting point is that inside these curves we are able to note special patterns, that should be investigated with more detail (see Fig.~\ref{fig:8}). 

\begin{figure}[t!]
\begin{center}
\hspace{0.7cm}\includegraphics[width=85mm]{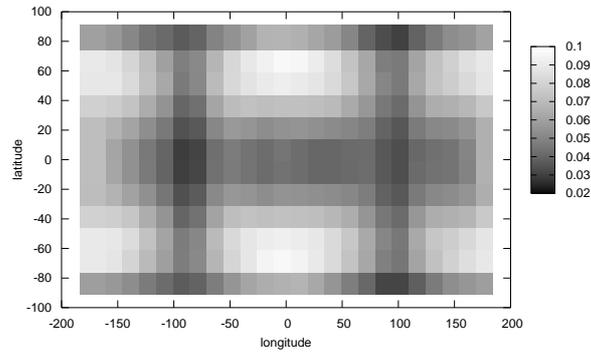}
\caption{\sf We show the density of impact caused by the dynamics associated with $\mathcal{W}^s(L_2)$ if the lunar craters distribution was uniform. We recall that the surface of the Moon is discretized in squares of $15^{\circ}\times 15^{\circ}$ and that the color bar refers to the number of impacts per unit of area normalized with respect to the total number of impacts found. The lighter the color the greater the impact density. The Earth -- Moon distance is assumed to be $d_{EM}=384400$ km.}
\label{fig:9}
\end{center}
\end{figure}

The density of impact on the Moon's surface produced by $\mathcal{W}^s(L_2)$ in this case is depicted in Fig.~\ref{fig:9}. The reader should be aware that we do not expect an apex concentration as before, due to the different collocation of points inside the $(y,\dot{y})$ and $(z,\dot{z})$ projections.

\begin{figure}[t!]
\begin{center}
\begin{tabular}{ccc}
\includegraphics[width=50mm]{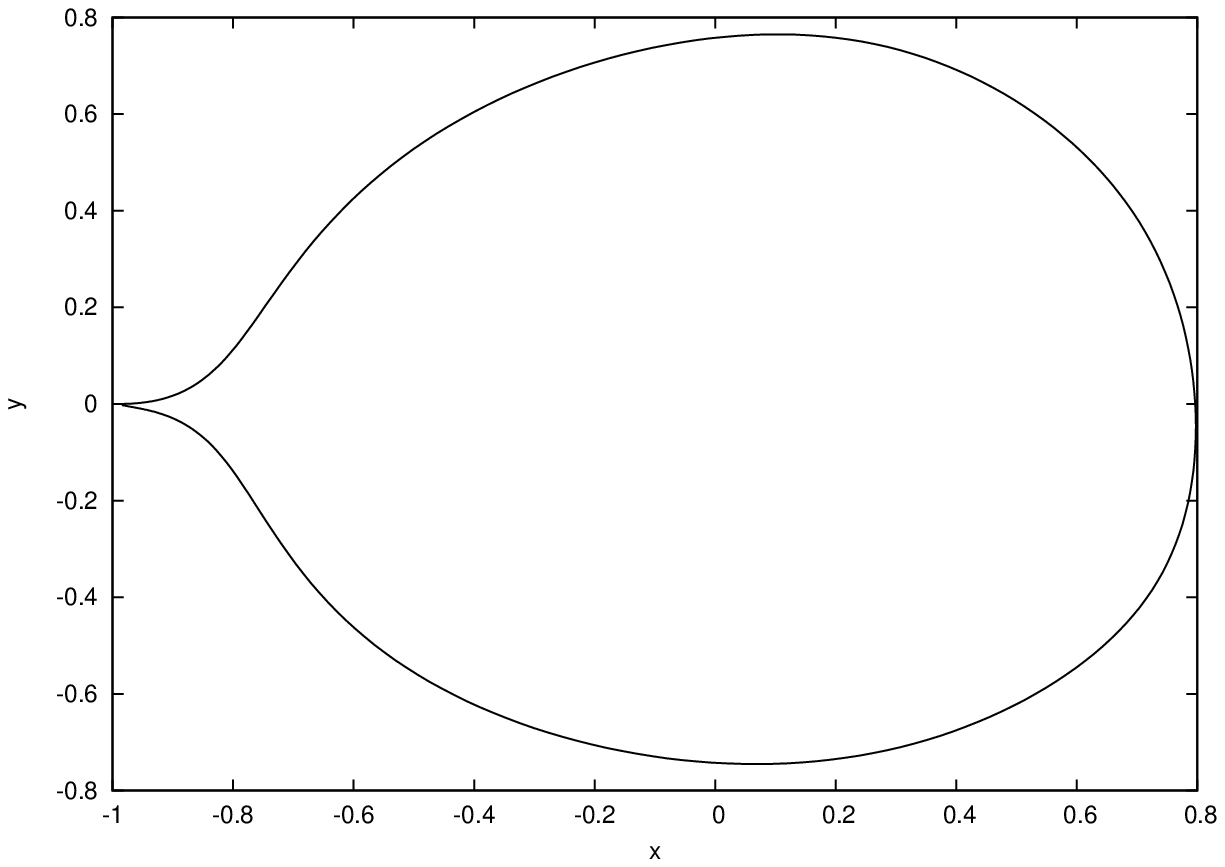}& 
\includegraphics[width=50mm]{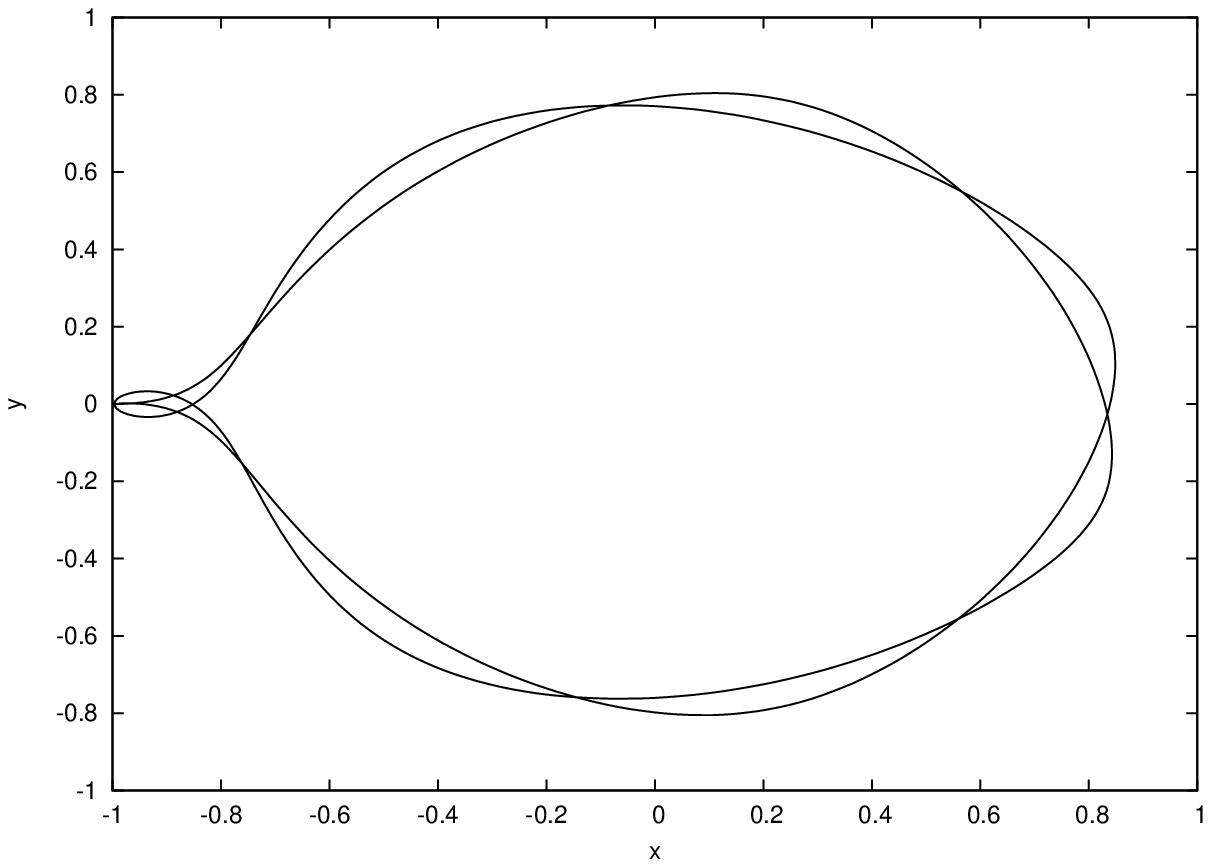}&
\includegraphics[width=50mm]{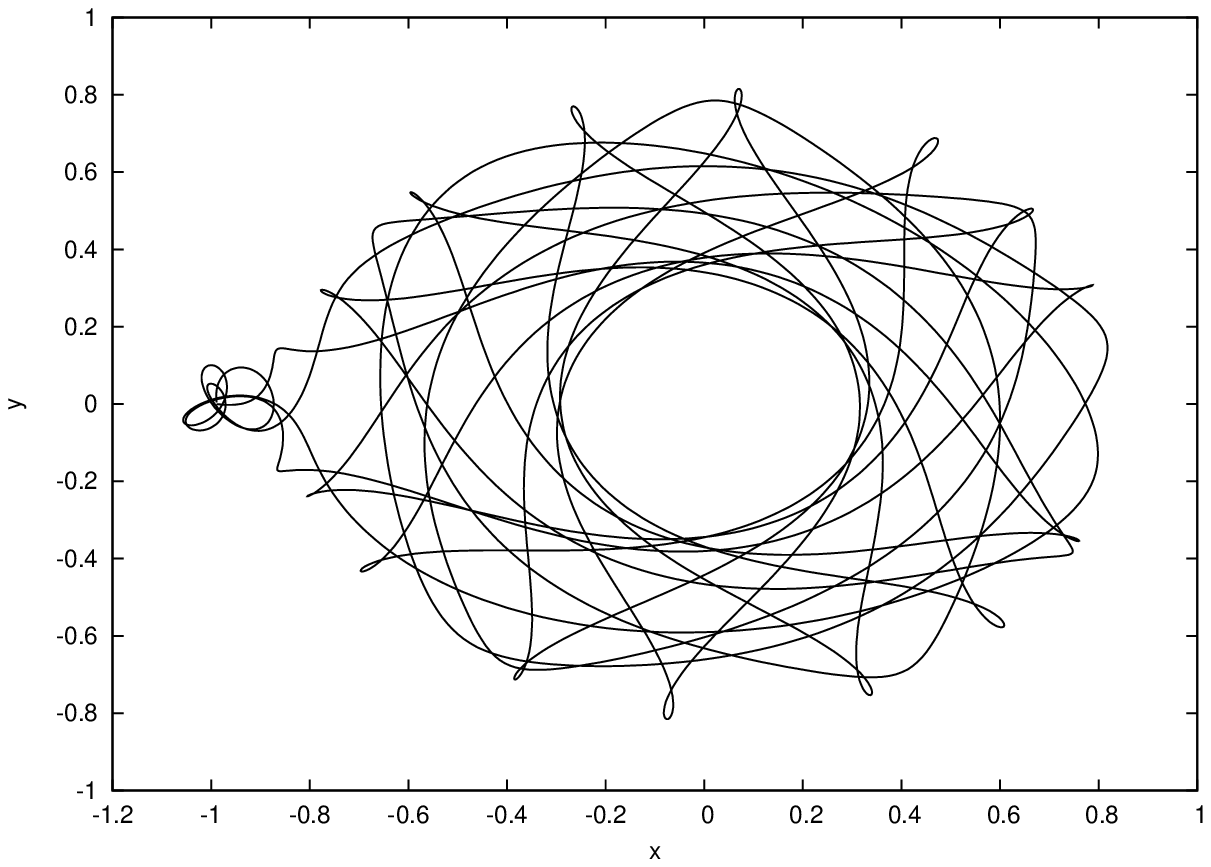}
\end{tabular}
\caption{\sf We show three low-energy trajectories departing from the surface the Moon and arriving there after revolving around the Earth. They have been derived assuming an uniform density of lunar impacts.}
\label{fig:10}
\end{center}
\end{figure}

On the other hand, there exist orbits that depart from the Moon with about the lunar escape velocity and return there with the same speed (see Fig.~\ref{fig:10}). They can travel along different paths, turning around either the Earth or the Moon one or several times. As explanation, we can hypothesize ejecta deriving from high-energy collisions. Such effect has already been predicted by other authors \cite{GBDLL}.

\section{The Bicircular Restricted Four -- Body Problem}\label{sec:5}

The Bicircular Restricted Four -- Body Problem \cite{CRR} considers the infinitesimal mass $P$ to be affected by the gravitational attractions of three primaries. We introduce this model in order to include the effect of the Sun in the Earth -- Moon system. The Earth and the Moon revolve in circular orbits around their common center of mass and, at the same time, this barycenter and the Sun move on circular orbits around the center of mass of the Earth -- Moon -- Sun system. 

The usual framework to deal with is the synodical reference system centered at the Earth -- Moon barycenter: in this way the Earth and the Moon are fixed on the $x$--axis as before and the Sun is supposed turning clockwise around the origin. We notice that the three massive bodies are assumed to move in the same plane and that the model is not coherent in the sense that the primaries motions do not satisfy the Newton's equations.

Let us take adimensional units as in the CR3BP and let $m_S=328900.5614$ be the mass of the Sun, $a_S$ be the distance between the Earth -- Moon barycenter and the Sun, $\omega$ be the mean angular velocity of the Sun in synodical coordinates and $\theta_0$ be the value associated with the rotation of the Sun with respect to the Earth -- Moon baycenter at $t=0$. If $\theta=\omega t$, then the position of the Sun is described by
\begin{eqnarray}
x_S&=& a_S\cos{(\theta-\theta_0)},\\
\label{eq:6}
y_S&=&-a_S\sin{(\theta-\theta_0)},\nonumber
\end{eqnarray}
and the equations of motion for the particle $P$ can be written as
\begin{eqnarray}
\ddot x-2\dot y&=& ~x-\frac{(1-\mu)}{r_{1}^{3}}(x-\mu)-\frac{\mu }{r_{2}^{3}}(x+1-\mu)-(x-x_S)\frac{m_S}{r_S^3}-\cos{(\theta-\theta_0)}\frac{m_S}{a_S^2},
\nonumber \\
\label{eq:7}
\ddot y+2\dot x&=& ~y-\frac{(1-\mu)}{r_{1}^{3}}y-\frac{\mu }{r_{2}^{3}}y-(y-y_S)\frac{m_S}{r_S^3}-\sin{(\theta-\theta_0)}\frac{m_S}{a_S^2},\\
\ddot z&=&~-\frac{(1-\mu)}{r_{1}^{3}}z-\frac{\mu}{r_{2}^{3}}z-z\frac{m_S}{r_S^3},\nonumber
\end{eqnarray}
where $\mu$ has the same meaning and value as the one introduced in Sec. \ref{sec:2} and $r_{1}= [(x-\mu)^{2}+y^{2}+z^{2}]^{\frac{1}{2}}$, $r_{2}=[(x+1-\mu)^{2}+y^{2}+z^{2}]^{\frac{1}{2}}$, $r_S=[(x-x_S)^{2}+(y-y_S)^{2}+z^{2}]^{\frac{1}{2}}$ are the distances from
$P$ to Earth, Moon and Sun, respectively.

We recall that this problem does not admit either first integrals or equilibrium points.

We note that also in the case of a planet without a moon, it is still possible to apply the BR4BP by considering two planets and the Sun. For instance, we can assume the Sun and Mercury to move as in the CR3BP and Venus to move around their barycenter on a circular orbit lying on the same plane. For more details you can refer to \cite{GG}.

\subsection{Results}\label{sec:5.1}

\begin{table}[h!]\label{tab:2}
\begin{center}
\caption{\sf We show the percentages of Moon's and Earth's impact found, for different values of Earth -- Moon distance $d_{EM}$ and initial phase of the Sun $\theta_0$.}
\begin{tabular}{c c c c}
\hline
\hline
$\mathbf{d_{EM}} (km)$ & $\mathbf{\theta_0}$ & $\mathbf{\%}$ {\bf Moon impacts} & $\mathbf{\%}$ {\bf Earth impacts}\\
\hline
\hline\\
\hline
\hline
232400 &36$^{\circ}$& 22.0 & 2.7\\
\hline
232400 &108$^{\circ}$& 13.9 & 3.3 \\
\hline
232400 &180$^{\circ}$& 14.4 & 3.0\\
\hline
232400 &252$^{\circ}$& 21.7 & 2.6\\
\hline
232400 &324$^{\circ}$& 10.5 & 4.9\\
\hline
\hline\\
\hline
\hline
270400 &36$^{\circ}$& 20.0 & 2.8\\
\hline
270400 &108$^{\circ}$& 10.1 & 3.7\\
\hline
270400 &180$^{\circ}$& 10.5 & 2.9\\
\hline
270400 &252$^{\circ}$& 20.1 & 2.2\\
\hline
270400 &324$^{\circ}$& 6.8 & 4.2\\
\hline
\hline\\
\hline
\hline
308400 &36$^{\circ}$& 17.0 & 2.4\\
\hline
308400 &108$^{\circ}$& 7.3 & 3.7\\
\hline
308400 &180$^{\circ}$& 8.1 & 2.1\\
\hline
308400 &252$^{\circ}$& 18.7 & 2.0\\
\hline
308400 &324$^{\circ}$& 4.2 & 2.7\\
\hline
\hline\\
\hline
\hline
384400 &36$^{\circ}$& 13.3 & 2.2\\
\hline
384400 &108$^{\circ}$ & 3.2 & 2.9\\
\hline
384400 &180$^{\circ}$& 3.9 & 1.3 \\
\hline
384400 &252$^{\circ}$& 14.8 & 2.1\\
\hline
384400 &324$^{\circ}$& 1.2 & 1.1\\
\hline
\hline
\end{tabular}
\end{center}
\end{table}
Now, our objective is to clarify if the effect of the Sun can spoil relatively the heavier concentration of impact on the leading side of the Moon found previously. Hence, we apply the BR4BP equations of motion to the same initial conditions considered within the CR3BP framework. Also in this case, we are able to attribute to $d_{EM}$ some specific values, which account for the rate of recession of the Moon with respect to the Earth. We notice that $a_S$ and $\omega$ change accordingly to $d_{EM}$, as we assume the adimensional set of units defined in Sec.~\ref{sec:3}. The simulation is carried on as in Sec.~\ref{sec:3}, apart from the fact that now we have to explore the behaviour corresponding to different $\theta_0$ (we take 5) and that we have to take care of impacts on the surface of the Earth. Finally, the maximum time span we allow to give birth to a lunar collision is of about 5 years. This choice is essentially due to the increasing computational effort.

\begin{figure}[b!]
\begin{center}
\begin{tabular}{ccc}
\includegraphics[width=50mm]{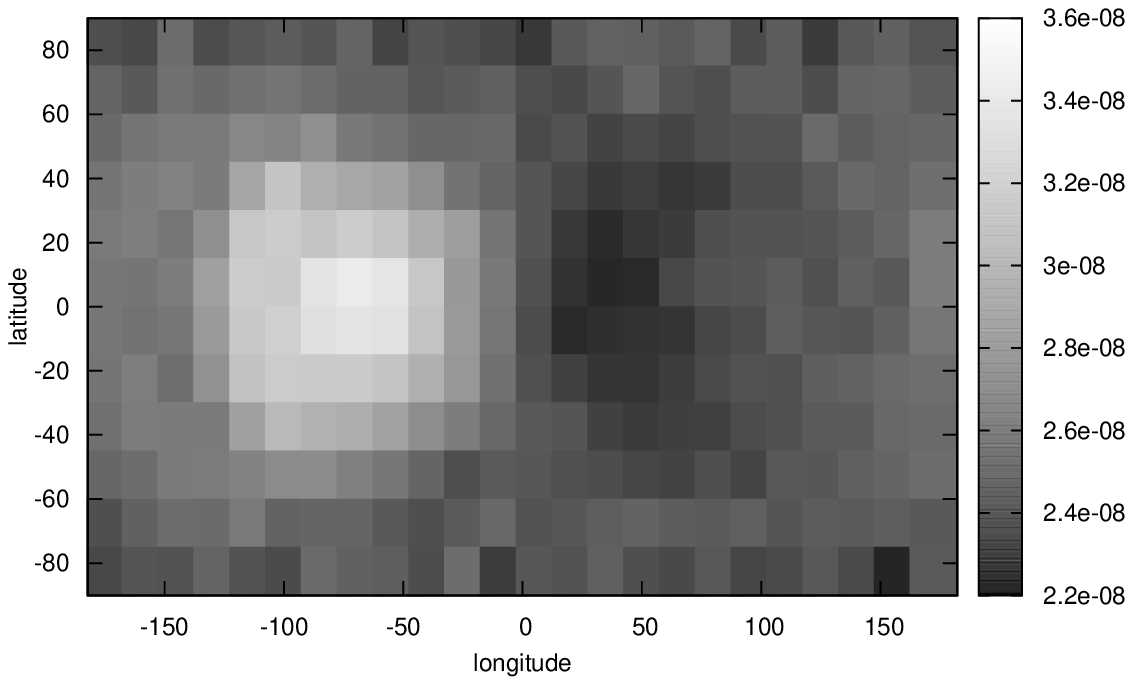}& 
\includegraphics[width=50mm]{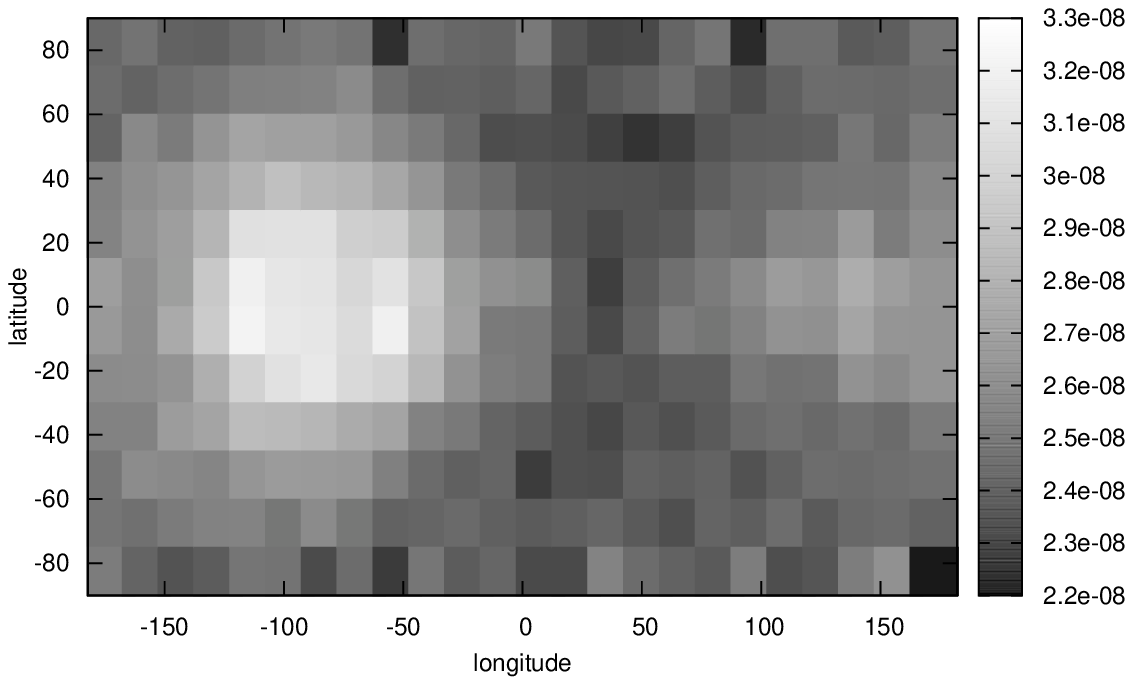}&
\includegraphics[width=50mm]{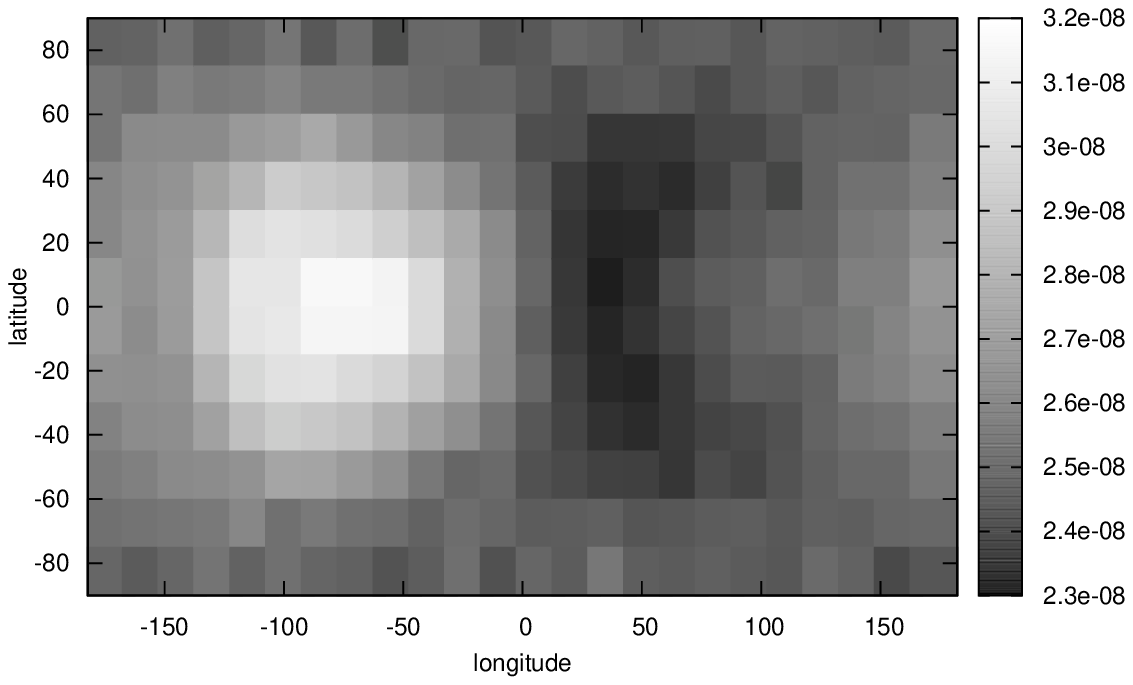}\\
\includegraphics[width=50mm]{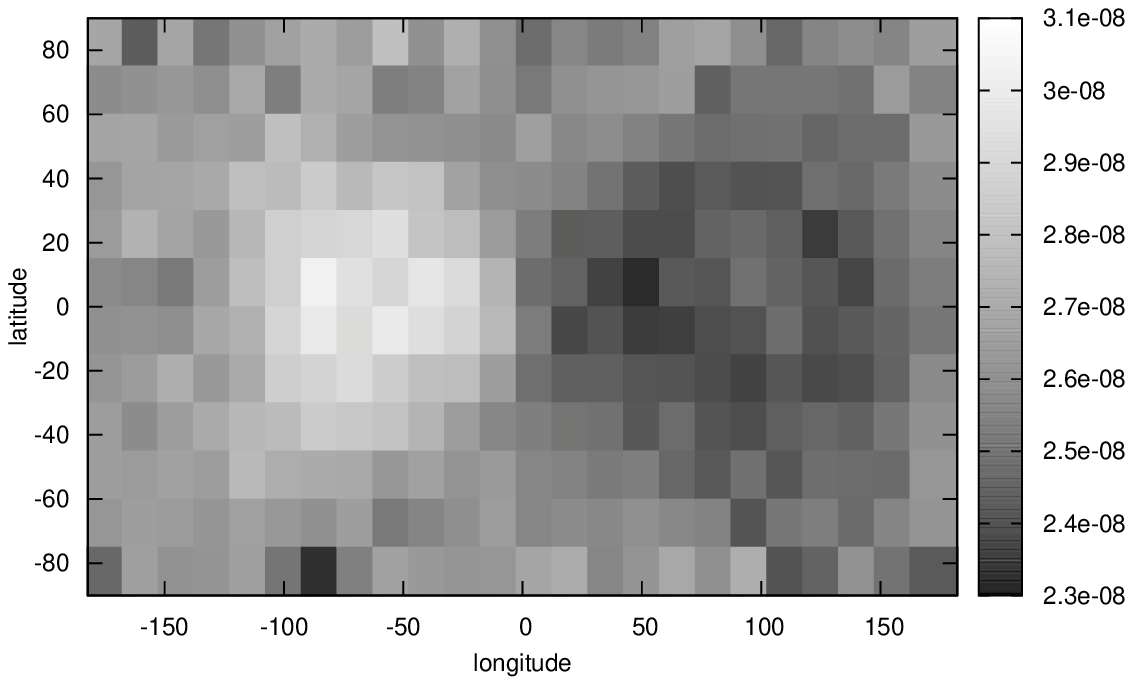}& 
\includegraphics[width=50mm]{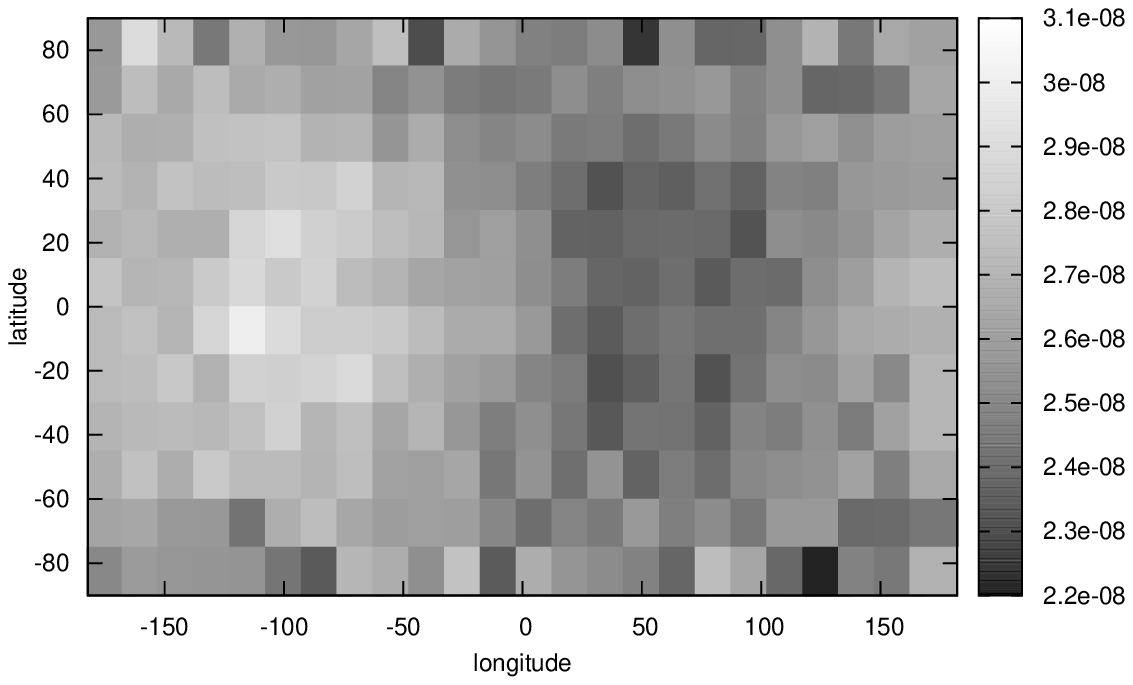}&
\includegraphics[width=50mm]{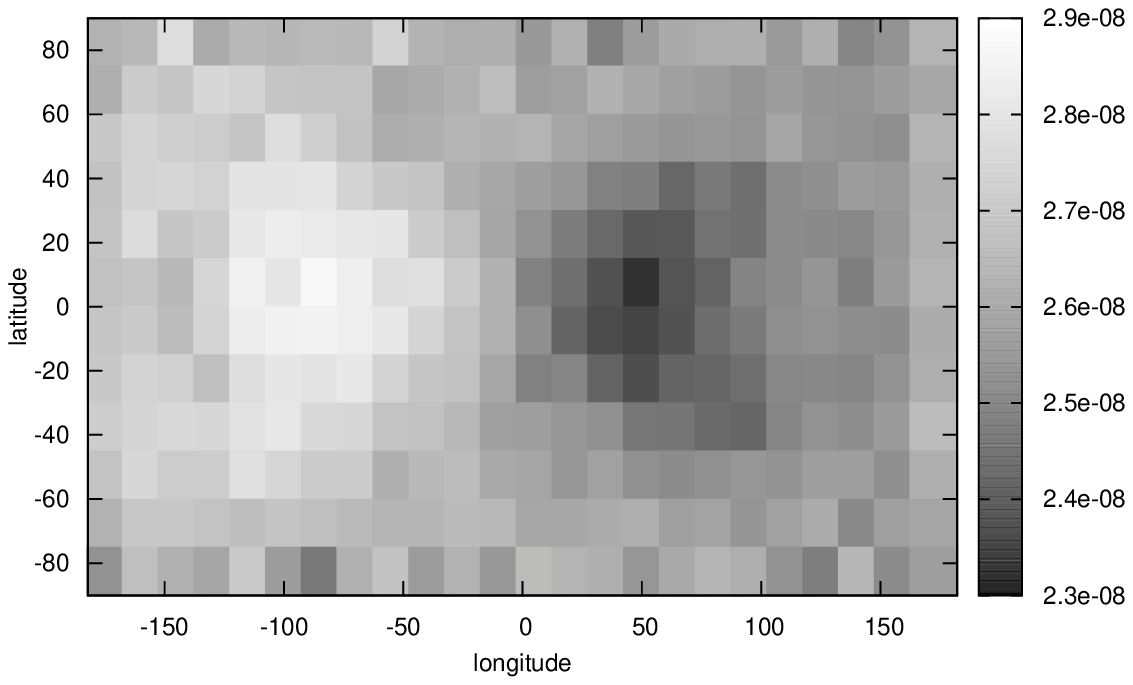}
\end{tabular}
\caption{\sf We display the density of impact (number of impacts per unit of area normalized with respect to the total number of impacts obtained) computed with the BR4BP equations of motion. The surface of the Moon is discretized in squares of $15^{\circ}\times 15^{\circ}$. The color bar indicates that the lighter the color of the square the greater the impact density. On the left, $\theta_0=36^{\circ}$; on the middle, $\theta_0=180^{\circ}$ and on the right we display the distribution due to all the five values (see Tab.~\ref{tab:2}) of $\theta_0$ considered.  On the top, $d_{EM}=232400$ km; on the bottom, $d_{EM}=308400$ km.}
\label{fig:11}
\end{center}
\end{figure}

Though we are aware that a deeper analysis using longer time intervales should be performed, significant results have already been obtained. They can be summarized as follows:
\begin{itemize}
\item[(a)] the percentage of impact depends on $d_{EM}$ and on the initial phase of the Sun, $\theta_0$: this is displayed in Tab.~\ref{tab:2};
\item[(b)] some trajectories collide with the Earth, the corresponding percentage is also shown in Tab.~\ref{tab:2};
\item[(c)] looking to Tab.~\ref{tab:2}, it is also clear that there exist values of $\theta_0$ more favorable to yield impacts with the Moon, while the Earth's percentage of impact has an almost constant trend;
\item[(d)] it looks like the relative Earth -- Moon and Earth -- Moon -- Sun distances, as well as the adimensional diameter of the Moon play a significant role in what concerns with the region of heavier lunar impact. In particular, the leading side collision concentration becomes more and more evident as $d_{EM}$ decreases.
\end{itemize}

In Fig.~\ref{fig:11}, we show the density of impact obtained when $d_{EM}=232400$ and $d_{EM}=308400$ km, respectively. For these plots, we consider $\theta_0=36^{\circ}$, $\theta_0=180^{\circ}$ and the distribution deriving from all the values of $\theta_0$ evaluated.

\section{A note on the computational effort}\label{sec:6}

As final remark, we note that all the simulation performed are quite expensive from a computational point of view, but that it was not an objective of this work to implement optimal programs. We use the UPC Applied Math cluster system, which consists in 26 Dell PowerEdge SC1425 servers, each with two Intel Xeon 3.2 GHz processors and 2 GB of RAM. In particular, we carry on parallel computations, using a node for each level of energy considered. In this way, the full computations for the CR3BP case take about 6 hours, for the BR4BP one about 1 day and the computations described in Sec.~\ref{sec:4} about 1 hour and 50 minutes. In all the numerical integrations, we adopt a 7-8 Runge-Kutta-Fehlberg method.

\section{Conclusions and future developments}\label{sec:7}

The main purpose of this paper is to establish a relationship between low-energy trajectories in the Earth -- Moon system and lunar impact craters. This is actually a quite wide and challenging topic, which involves knowledge related to mathematics, astronomy and geology. 

As primary goal, we define some tools which are effective in the determination of the dynamics pushing a massless particle under low-energy regimes. We exploit invariant objects within the Circular Restricted Three -- Body Problem approximation, in particular transit trajectories lying inside the stable invariant manifold associated with the central invariant manifold of the $L_2$ equilibrium point. We implement a method that allows to reproduce the behaviour associated with the unstable component of any central orbit and does not need to distinguish between them. This fits with our investigation, because we are interested in minor bodies collisions that take advantage of the channels represented by the whole hyperbolic manifolds. To this purpose, we adopt well-known procedures to compute periodic Lyapunov orbits together with their hyperbolic invariant manifolds. 

With this approach, we perform extensive numerical simulations to determine both the lunar region of heavier impact and the sources of a potential uniform craters distribution. We also look for the influence of the Sun on these paths, by means of the Bicircular Restricted Four -- Body Problem. Several outcomes can be highlighted, even if they have to be seen as patterns that require a more robust proof: further calculations with different dynamical and astronomical models are in progress. The investigation carried out is promising from many points of view, as it indicates future developments that are worth to be considered.

Without the effect of the Sun, we get a confirmation that the neighbourhood of the apex of the surface of the Moon is the region where most collisions take place. We remark that the impact trajectories simulated reach the surface of the Moon with the lowest possible velocity: this point does not corrupt the apex concentration that other authors discovered without this restriction. The total time span considered (60 years) is sufficient to describe the general behaviour of the massless particles and no distinctions among different values for the Earth -- Moon distance were observed.

However, the gravitational force exerted by the Sun seems to blur the above phenomenon. Changing the ratio between the Earth -- Moon -- Sun distance and the Earth -- Moon one, we notice different patterns. From our computations, it turns out that in more ancient epochs the low-energy lunar impacts were focused on the Moon leading side, but this is not true going further in time. Moreover, we get evidence that the position of the Sun with respect to the Earth -- Moon barycenter affects the distribution of lunar impacts. These are the first aspects we plan to study with more detail. For instance, it would be deserving to integrate the BR4BP equations of motion for a longer interval of time and for a greater number of values of the initial phase of the Sun to understand the real nature of these numerical observations. 

On the other hand, we realize that small craters can also be generated by the impact of dust arising from more energetic collisions than the ones investigated here. Such phenomenon comes from the existence of periodic orbits that cross the surface of the Moon, that is, double collision orbits. In the future, we would like to see how they are transformed by the perturbation of the Sun.

A natural step would be to add the gravitational attractions of other planets to see their consequences on the orbits simulated. This will be done by means of a Restricted n -- Body Problem, using position and velocity of the primaries given by the JPL ephemerides (for instance the DE405 ones) and taking several initial epochs to compare the whole outcome.

Moreover, we would like to link our methodology with real observational data, concerning either the existing lunar craters and the orbital parameters at a certain epoch of a given set of Near Earth Objects. This information would affect especially the way we generate the initial conditions corresponding to transit orbits.

Finally, to apply the same kind of analysis to the terrestrial planets would be of large interest. Starting from the CR3BP approximation, we mean to study the density of impact provided by Sun -- planet low-enery orbits and then to add further gravitational effects, trying to figure out the orbital elements and also the regions in the phase space which more likely lead to collision.

\begin{acknowledgements}
This work has been supported by the Spanish grants MTM2006--05849 (E.M.A., G.G.), MTM2009--06973 and 2009SGR859 (J.J.M.) and by the Astronet Marie Curie fellowship (E.M.A.). We also acknowledge the use of the UPC Applied Math cluster system for research computing (see http://www.ma1.upc.edu/eixam/index.html).
\end{acknowledgements}

\begin{appendix}
\section*{Appendix: Knuth Shuffle Algorithm}

In this work, to obtain a sequence of random real numbers uniformly distributed between 0 and 1 $\langle U_n\rangle$, we exploit the following linear congruential method. Let $m=2^{31}-1=2147483647$, $a=7^5=16807$, $q=127773$ and $r=2836$, then
\begin{eqnarray*}
T_k&=&a(X_k\textrm{mod}~q)-r[X_k/q],\\
aX_k\textrm{mod}~m&=& \begin{cases}
              \begin{array}{ll}
                   T_k ~~~~~~~~~~~~~~~\textrm{if}~~~~~~T_k\ge 0\\
                   T_k+m ~~~~~~\textrm{if}~~~~~~T_k < 0\\
              \end{array}
       \end{cases},\\
X_{k+1}&=&aX_k\textrm{mod}~m,\\
U_{k+1}&=&X_{k+1}/m.
\end{eqnarray*}

As shuffle algorithm, we mean a procedure which aims at reorder a given sequence $\langle U_n\rangle$ to improve its quality. We adopt this approach:
\begin{enumerate}
\item we initialize an auxiliary sequence $\langle V_0,V_1,\dots, V_p \rangle$ with the first $p$ values of the $X-$sequence;
\item we define $Y=X_{p+1}$;
\item $j=\lfloor pY/m \rfloor$;
\item $Y=V_j$;
\item $V_j=X_{p+1}$;
\item the final output is represented by $Y$.
\end{enumerate}

For further details, please refer to \cite{K}.
\end{appendix}

\end{document}